\documentclass[lettersize,journal]{IEEEtran}
\usepackage{amsmath,amsfonts,amssymb}
\usepackage{algorithm,algpseudocode}
\algnewcommand{\algorithmicforeach}{\textbf{for each}}
\algdef{SE}[FOR]{ForEach}{EndForEach}[1]
  {\algorithmicforeach\ #1\ \algorithmicdo}%
  {\algorithmicend\ \algorithmicforeach}%
\usepackage{enumerate}
\usepackage{algpseudocode}
\usepackage{array}
\newcolumntype{?}{!{\vrule width 1pt}}

\usepackage[caption=false, labelformat=parens, subrefformat=parens]{subfig}
\usepackage{textcomp}
\usepackage{stfloats}
\usepackage{url}
\usepackage{diagbox}
\usepackage{verbatim}
\usepackage{graphicx}
\usepackage{etoolbox}
\usepackage{overpic}
\usepackage{cite}
\usepackage{booktabs}
\usepackage[hidelinks]{hyperref} 
\usepackage{makecell}
\usepackage[export]{adjustbox}
\usepackage{array, multirow}
\usepackage{tabularx}
\usepackage[switch]{lineno}

\ifdefined\pdfinfoomitdate
\fi
\hypersetup{pdfauthor={},pdftitle={},pdfsubject={},pdfkeywords={},pdfcreator={},pdfproducer={}}
\begin{document}

\title{Physics-Constrained Deep Learning Model for Contactless Blood Pressure Monitoring from Triaxial Bodyseismography}
\author{Yuanyuan~Zhang, Yida~Zhang, Jiahui~Li, Yuyan~Wu, Fei~Dou, Xiao~Yin,\\Zhenlin An, Hae Young Noh, WenZhan~Song, \IEEEmembership{Senior~Member,~IEEE}
\thanks{This research has been approved by the Institutional Review Board of University of Georgia with application number No.PROJECT00010221, and is also approved by Medical Ethics Committee of Yixing People's Hospital under application No.LS2024-KE-084-02.}
\thanks{Yuanyuan Zhang, Yida Zhang and WenZhan Song are with the School of Electrical and Computer Engineering, University of Georgia, Athens, 30603, United States, (email: yz52357@uga.edu, wsong@uga.edu).}
\thanks{Jiahui Li, Fei Dou and Zhenlin An are with the School of Computing, University of Georgia, Athens, 30603, United States.}
\thanks{Yuyan~Wu and Hae~Young~Noh are with the Department of Civil and Environmental Engineering, Stanford University, Stanford, 94305, United States.}
\thanks{Xiao Yin is with the Yixing People’s Hospital, Wuxi 214200, China.}
\thanks{\textit{Corresponding author: WenZhan Song}}}%

\maketitle

\begin{abstract}
Ballistocardiography (BCG) is promising for unobtrusive long-term blood pressure (BP) monitoring in laboratory settings, but traditional BCG signals are vulnerable to the variations in body-bed interaction with shifted fiducial points in temporal or amplitude axis, and BP varies with personal hemodynamic changes, causing misaligned representations that affect model generalizability and robustness. In this work, we propose Phy-BP, a physics-constrained BP estimation framework based on triaxial bodyseismography (BSG) acquired using bed-mounted sensors as an extension of single-axis BCG. Firstly, we design an adaptive quality-control algorithm combining neighboring beat patterns and universal cardiogenic templates to retain reliable cardiac components. Secondly, we propose a physical model of wave propagation through the body-bed system to constrain latent feature evolution, aligning triaxial representations generated by a shared cardiogenic excitation. These mechanical constraints connect multi-axis feature learning to body-bed dynamics and complement data-driven regression from vibration signals. Evaluation on a 162-hour hospital dataset from 21 subjects against invasive arterial BP yields a mean absolute error of 5.07 mmHg, a prediction-error standard deviation of 7.24 mmHg, and a Pearson correlation coefficient of 0.86 for mean arterial pressure. This framework is intended for unobtrusive monitoring during rest and sleep, with potential applications in overnight hospital and home monitoring.
\end{abstract}

\section{Introduction}
Accurate and long-term blood pressure (BP) monitoring is important for capturing nocturnal and longitudinal hemodynamic variations in both home and clinical settings, supporting sleep-related BP assessment, early cardiovascular risk detection and routine overnight monitoring~\cite{huai2024bp,huang2024ai}. Although invasive arterial blood pressure (ABP) monitoring remains the gold standard for continuous BP measurement, it requires catheterization and is associated with increased clinical burden and risks such as infection and thrombosis~\cite{litton2012picco,zhang2025blood}. As a non-invasive alternative, cuff-based devices are widely used for intermittent BP measurement, but their accuracy can be affected by factors such as device algorithms, cuff size, and arm position, while frequent cuff inflation may also cause discomfort, sleep disturbance, and reduced compliance during long-term use~\cite{li2025practicalbp,shahrbabak2025silico}.

Recent studies explore cuffless BP monitoring using non-invasive physiological signals. For instance, photoplethysmography (PPG) and seismocardiography (SCG) have been used to estimate BP through pulse transit time (PTT) or cardiac activity intensity~\cite{li2025practicalbp,chen2026intrinsic,wang2025pitn,wang2024accurate,chang2019cuff,wu2025ubicon}. However, these contact-based approaches require sustained skin contact and are therefore unsuitable for patients with skin damage or for long-term/overnight monitoring~\cite{zhang2023overview}. To achieve more unobtrusive monitoring, camera/radar-based methods have been explored for remote physiological sensing, but camera systems raise privacy concerns, while current radar methods still suffer from limited signal-to-noise ratio (SNR) and short monitoring range (e.g., within $0.7$ meters) under practical conditions~\cite{wu2022facial,zhang2024radarODE,zhang2025high,cao2025mmwave,shi2025mmbp}.

\begin{figure*}[tb] 
    \centering 
    \includegraphics[width=2\columnwidth]{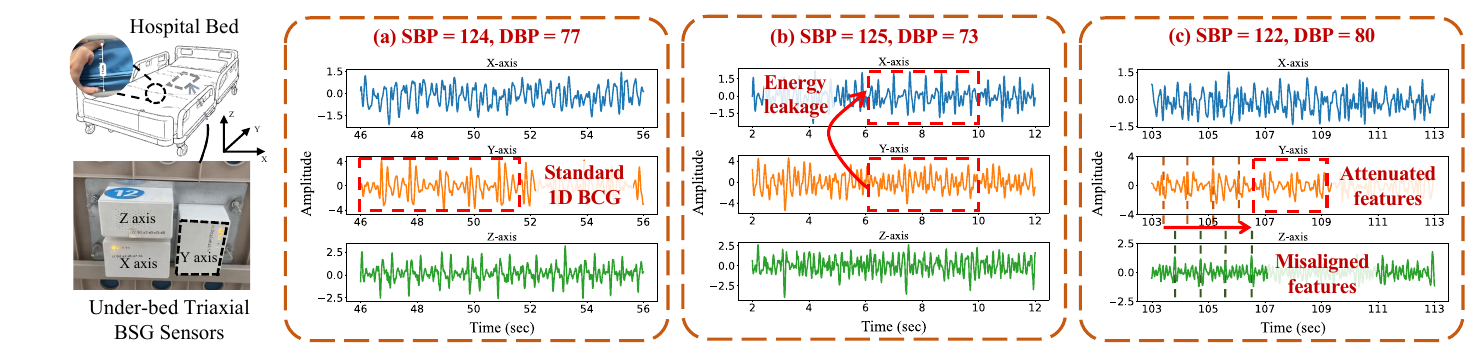}
    \caption{Illustration of BSG sensors placement and signals under similar systolic blood pressure (SBP) and diastolic blood pressure (DBP): (a) Standard BSG signal with the majority of features shown in Y-axis; (b) Energy leakage to X-axis with no observable features in Y-axis; (c) Distorted BSG signal due to soft/thick mattress with attenuated and shifted fiducial peaks.}
    \label{fig:bsg_quality}
\end{figure*}

Compared with the aforementioned solutions, bed-mounted vibration sensors can detect the minor vibrations caused by heartbeat and body recoil along the foot-head axis (i.e., ballistocardiography (BCG))~\cite{huang2024ai,starr1955studies,starr1959studies}, as shown in Fig.~\ref{fig:bsg_quality}(a). Although the BCG measurements are inherently tied to the bed or other transmission media (e.g., chair~\cite{kim2018ballistocardiogram}), this fixed coupling can help maintain relatively stable signal quality and support continuous overnight monitoring~\cite{chen2025selfdenoiser}. High-quality BCG can be measured using various sensors (e.g., fiber-optic sensor~\cite{huang2024ai}, electromagnetic films~\cite{zhang2026bcg}, seismic sensor~\cite{chen2025selfdenoiser}), and numerous studies have demonstrated that the BCG signals contain cardiovascular information for downstream applications, such as sleep status monitoring~\cite{li2020smart} and arrhythmia detection~\cite{lyu2025contactless}, but BP estimation requires fine-grained cardiac features other than respiration and heart rates.

Existing studies on BCG-based BP estimation can be categorized into two paradigms. The first paradigm requires simultaneous collection of other vital signs (e.g., electrocardiogram (ECG)) to estimate PTT first, and is not purely contactless~\cite{kim2018ballistocardiogram}. In contrast, recent research leverages deep learning models to learn BP-related representations directly from raw BCG waveforms due to the powerful non-linear mapping abilities~\cite{xiong2025enhancing,chu2024vessel}, thereby alleviating reliance on other modalities~\cite{huang2024ai,zhang2026bcg,huai2024bp}. However, the existing work shares some major limitations:
\begin{itemize}
\item \textbf{Inaccurate BP ground truth measurement:} Most BP estimation studies use intermittent BP readings from oscillometric devices as ground truth for convenience, while such measurements cannot capture beat-to-beat BP dynamics and thus cannot provide continuous hemodynamic monitoring as a potential clinical reference~\cite{li2025practicalbp,huang2024ai}. In addition, the accuracy of cuff-based measurements is highly affected by cuff size/placement, arm posture, and physiological fluctuations during repeated inflation~\cite{liu2024arm,kallioinen2017sources}.
\item \textbf{Unfaithful cardiac features in 1D BCG:} Standard 1D BCG captures body recoil along the Y-axis (foot-head direction), and its I, J, and K peaks correlate with cardiac force~\cite{starr1959studies}. However, our measurements show that 1D BCG misses part of the bed response to cardiac activity, as vibration energy can leak into the X- and Z-axes under different postures (Fig.~\ref{fig:bsg_quality}(b)). Changes in body-bed interaction due to mattress and body weight can also attenuate BCG features (Fig.~\ref{fig:bsg_quality}(c)).
\item \textbf{Reliance on purely data-driven methods:} Although studies have linked BCG fiducial points to cardiac force~\cite{huang2024ai,kim2018ballistocardiogram,starr1959studies}, these relations are not explicitly modeled for BP estimation based on cardiogenic excitation and wave propagation theory. Most existing work instead trains black-box deep learning models on BCG-BP pairs. Uncertainties in BP ground truth and BCG morphology can misalign extracted representations, limiting model generalizability and robustness~\cite{wu2026noncontact}.
\end{itemize}

In this study, we leverage triaxial bodyseismography (BSG)~\cite{song2025multi} measured by seismic sensors (Fig.~\ref{fig:bsg_quality}), as an extension of the 1D BCG signal for BP estimation to faithfully reveal the cardiac features. The triaxial BSG signals are firstly mapped through nonlinear encoders into high-dimensional representations. The feature representations are then projected into latent states to capture the underlying body-bed dynamics. To govern the temporal evolution of the latent states and promote cross-axis alignment, we model 3D wave propagation through the body-bed system and derive a physical constraint encouraging approximately linear dynamics under a shared cardiogenic excitation. Studies in other fields have shown that incorporating physical knowledge into deep learning can improve predictive accuracy and robustness~\cite{bose2024invariance,roy2023elasticity}. In addition, the invasive ABP waveform is used as a gold-standard reference for the model training and evaluation, along with simultaneously collected ECG and PPG signals for a fair and comprehensive comparison. The contributions of this work are listed as follows:
\begin{itemize}
\item To the best of our knowledge, this is the first study to explicitly characterize how cardiogenic excitation propagates through the body-bed coupling system. We establish a physical model based on wave propagation theory to interpret how the measured triaxial BSG signals are jointly shaped by cardiogenic driving forces and the dynamic response of the body-bed system.
\item A BSG-based BP estimation framework, Phy-BP, is proposed with an adaptive quality-control module that filters out low-SNR signals by considering both neighboring beat patterns and universal cardiogenic templates, ensuring enriched and faithful cardiogenic features for further BP estimation.
\item Phy-BP also contains a deep learning model for physics-constrained representation learning, and a dynamical system derived from the body-bed system governs the temporal evolution of learned latent states. The axis-specific states are encouraged to follow the same dynamic system under a common cardiogenic excitation, providing a physical basis for cross-axis feature alignment.
\item Using invasive ABP measurements as the reliable reference, the proposed Phy-BP was validated on a $162$-hour hospital dataset collected from $21$ patients. The results of extensive experiments demonstrate that Phy-BP outperforms existing BCG- and PPG/ECG-based BP estimation methods, achieving a $15.21\%$ overall gain relative to the best-performing model.
\end{itemize}

The paper proceeds as follows. Section II introduces background and challenges, and Section III presents methodology. Section IV describes the experiments and dataset. Section V reports results, and Section VI concludes this paper.

\section{Background and Challenges}\label{sec:bg}
\subsection{Physiological Rationale for BSG-based BP Monitoring}
From a hemodynamic perspective, BP is jointly determined by stroke volume (SV), heart rate (HR) and total peripheral resistance (TPR) as:
\begin{equation}\label{equ:tpr}
\text{BP} \propto \text{CO}\times\text{TPR} = \text{SV}\times \text{HR}\times\text{TPR},
\end{equation}
with SV and HR together determining the amount of blood pumped into the arterial system (i.e., cardiac output (CO)), and TPR characterizing the resistance imposed by the peripheral vasculature to blood flow. During each cardiac cycle, ventricular contraction ejects blood into the arterial tree and generates pulsatile changes in pressure and blood momentum, inherently linking the measurements from the arterial system (i.e., BP) and the cardiac mechanical system (i.e., BSG).

Specifically, BSG does not directly measure arterial pressure but captures the subtle body-bed recoil induced by blood acceleration, and the measured BSG waveform can faithfully reveal the cardiac force and is regarded as an indirect mechanical projection of the underlying hemodynamic process~\cite{starr1959studies}. However, the absolute BP level is further modulated by subject-specific hemodynamic properties (i.e., TPR), and the instantaneous TPR condition cannot be directly observed from non-invasive physiological signals. Therefore, all the cuffless or non-invasive BP estimation methods require subject-specific calibration at the initial use stage, and periodic recalibration is also necessary to compensate for physiological drift and changes in hemodynamic state~\cite{li2025practicalbp,huang2024ai,zhang2026bcg}.

\begin{figure}[tb] 
    \centering 
    \includegraphics[width=0.8\columnwidth]{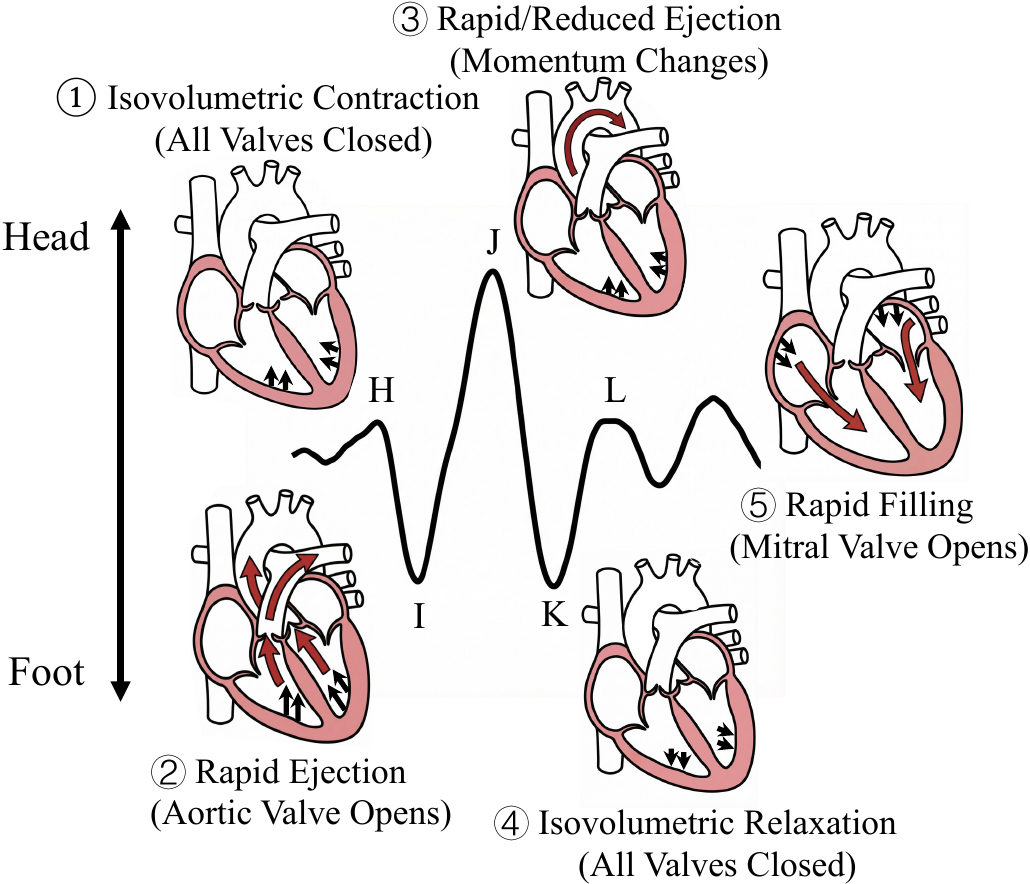}
    \caption{Correspondence between fiducial points and cardiac events as summarized in~\cite{starr1950standardization,starr1955studies,starr1959studies}.}
    \label{fig:template} 
\end{figure}

In contrast, the relationship between CO and BSG fiducial points was validated and explained through a series of necropsy studies in the 1950s, and the cardiogenic origins of these fiducial points are illustrated in Fig.~\ref{fig:template}. The H, I, and J waves correspond to isovolumetric contraction, aortic valve opening, and the subsequent acceleration of blood through the ascending aorta and aortic arch, respectively, with the J wave reflecting the largest change in blood momentum. The K and L waves are associated with blood-flow deceleration caused by peripheral resistance and arterial elasticity and with rapid ventricular filling during early diastole, respectively.

\subsection{Advantages and Challenges of BSG-based BP Monitoring}
The physiological relationship described in the last subsection encourages the development of BP estimation algorithms based on 1D BCG signals\cite{kim2018ballistocardiogram,huang2024ai}, but a single axis is not sufficient to capture the faithful body recoil induced by cardiogenic excitation, as shown in Fig.~\ref{fig:bsg_quality}(b) and Fig.~\ref{fig:bsg_quality}(c). To faithfully capture cardiac activity through bed vibration signals, our research mounted three seismic sensors under the bed frame with different orientations for 3-axis sensing and can capture standard BCG signals in the Y-axis with the leaked cardiac features in X- or Z-axis, as shown in Fig.~\ref{fig:bsg_quality}(a) and Fig.~\ref{fig:bsg_quality}(b).

Although it is natural to directly use triaxial BSG signals as three independent input channels for deep learning models, existing deep learning models operate as black boxes and can only process 1D BCG signals. In addition, an associated challenge with triaxial BSG is that the features may shift because the soft and thick mattress acts as a lag filter, as shown in Fig.~\ref{fig:bsg_quality}(c). As a result, the physiological inconsistencies (e.g., peak attenuation/shift and inaccurate cuff-based BP ground truth) across samples and axes contaminate BP-related hemodynamic information and prevent deep learning models from identifying robust, invariant representations. In summary, we conclude the following challenges to effectively extract cardiac features from triaxial BSG signals for BP estimation:
\begin{enumerate}
\item Obtain faithful BP measurements that can serve as reliable supervisory signals for developing and validating contactless BP monitoring systems toward clinical use.
\item Establish a physical model for the body-bed system to describe how a single cardiogenic excitation propagates in a body-bed system and is observed as triaxial BSG signals across the three axes.
\item Integrate such physical dynamics into the deep learning model to align the triaxial features under a shared latent state evolution, improving robustness and generalizability under varying signal morphology and limited-data scenarios.
\end{enumerate}

\begin{figure*}[tb] 
    \centering 
    \includegraphics[width=1.8\columnwidth]{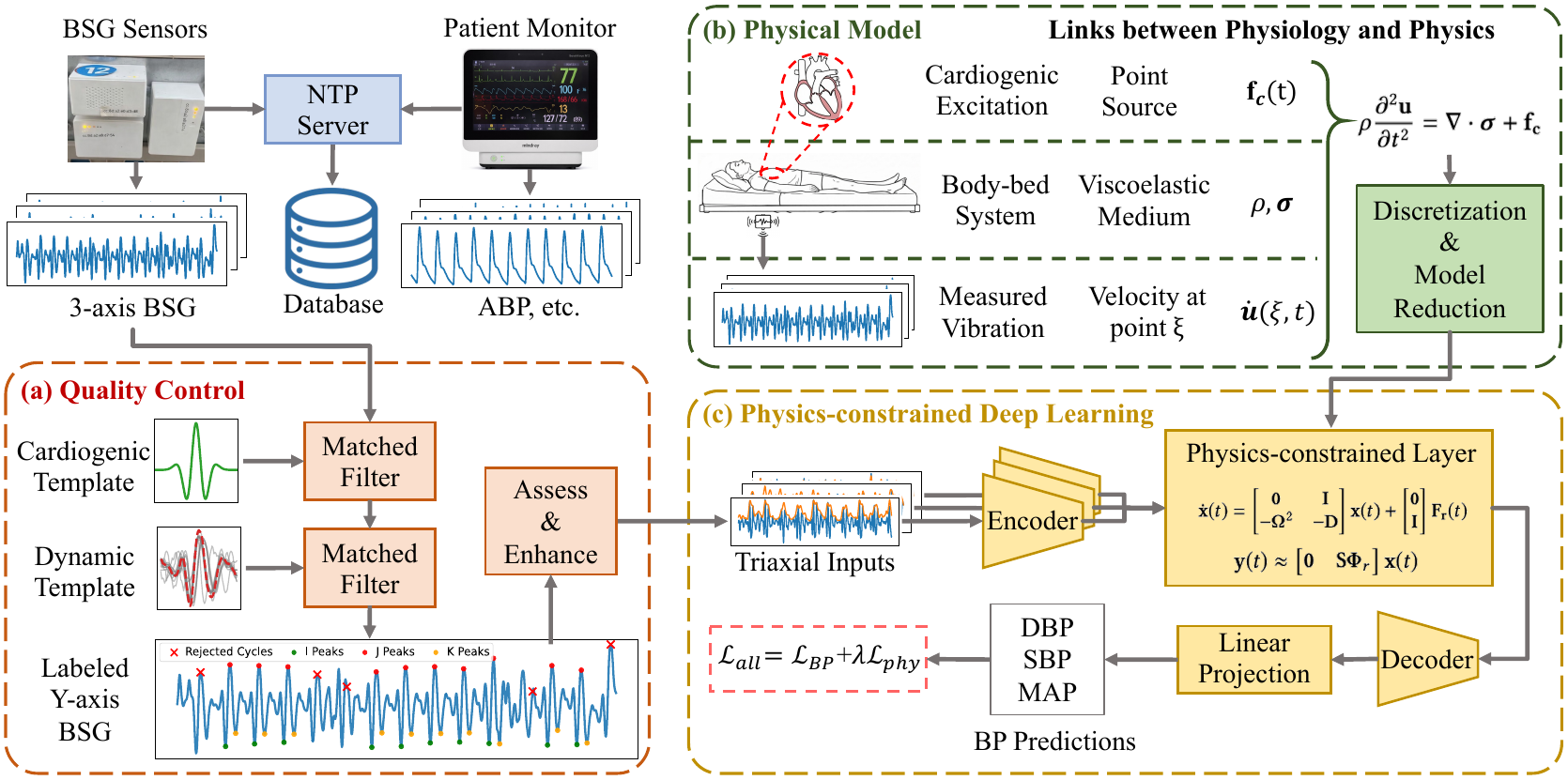}
    \caption{Overview of Phy-BP: (a) Quality control based on Y-axis BSG to retain signals rich in cardiogenic components; (b) Physical model that governs wave propagation excited by heartbeat; (c) Deep learning model designed with the triaxial feature extraction aligned and constrained by the physical model.}
    \label{fig:overview} 
\end{figure*}

\section{Methodology}\label{sec:method}
\subsection{Overview}
To overcome the aforementioned challenges, we propose Phy-BP to realize BSG-based BP estimation using deep learning, with a physical model established and embedded for cross-axis feature alignment. In addition, the data are collected from real patients in hospital settings to enable ethical ABP measurement as a gold standard for model training and validation. All data streams are timestamp-synchronized via a Network Time Protocol (NTP) server, therefore ensuring that heartbeat aligns with corresponding mechanical and hemodynamic responses across different vital signs. The Phy-BP pipeline is shown in Fig.~\ref{fig:overview} with three modules:
\begin{itemize}
\item \textbf{Quality control: }As a preprocessing step, all measured BSG will be automatically assessed by two matched filters to estimate SNR based on the cardiogenic components contained in the Y-axis. Eventually, the labeled Y-axis BSG signals can be easily assessed by the number of remaining single-cycle BSG patterns, ensuring effective deep learning training or inference.
\item \textbf{Physical model: }To assist the further feature alignment across 3 axes, we model the wave propagation in a viscoelastic medium (i.e., body-bed system) excited by a point source (i.e., heartbeat) using the partial differential equation (PDE), and the 3-axis BSG measurement is modeled as a single-point velocity observed under the bed, as shown in Fig.~\ref{fig:overview}(b).
\item \textbf{Physics-constrained deep learning model: }The general model design uses the traditional encoder-decoder architecture, but the extracted 3-axis features are aligned via a physics-constrained layer to govern the feature evolution, as shown in Fig.~\ref{fig:overview}(c). Specifically, the proposed layer aims to preserve the axis-specific characteristics arising from cardiogenic energy leakage across the three axes, and reduce feature inconsistency caused by waveform attenuation and temporal distortion.
\end{itemize}

\subsection{Quality Control}
Considering the fact that our dataset is collected from the real hospital scenario, a considerable portion of the recorded segments may be contaminated and lack sufficient cardiogenic content due to patient body movements and other uncontrollable disturbances in routine clinical environments. Therefore, quality control is essential to filter out low-quality segments and preserve only physiologically meaningful observations as shown in Fig.~\ref{fig:overview}(a).

\subsubsection{Matched Filter with Cardiogenic Template for Universal Beat Patterns}
In this work, the matched filter is adopted to quantify the similarity between an observed signal $s(t)$ and a predefined heartbeat template $\psi(t)$ through a convolution operation:
\begin{equation}\label{equ:match}
y_f(t) = (s * h)(t) = \int_{-\infty}^{\infty} s(\tau)\, h(t-\tau)\, d\tau,
\end{equation}
where $y_f(t)$ denotes the filter response, and the matched filter $h(t)$ is defined as the time-reversed version of the template signal as $\psi(-t)$.

\begin{figure*}[tb] 
    \centering 
    \includegraphics[width=1.8\columnwidth]{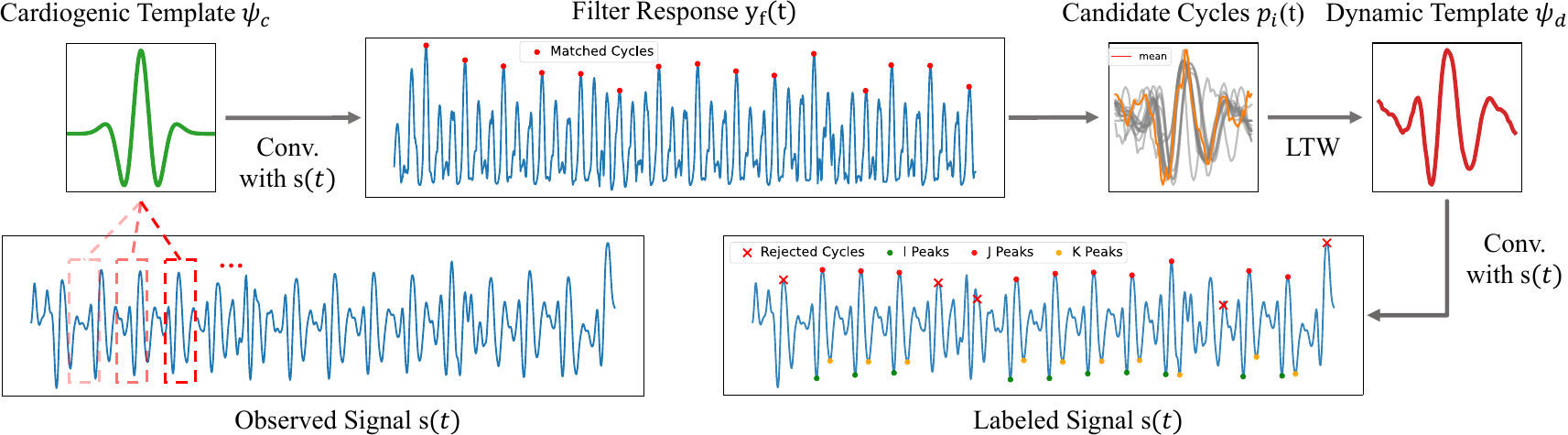}
    \caption{Design of matched filters with cardiogenic and dynamic templates.}
    \label{fig:match_filter} 
\end{figure*}

Considering the common pattern shared by Y-axis BSG signals with explicit physiological meanings as introduced in Fig.~\ref{fig:template}, we first design a universal template $\psi_c(t)$ that matches the morphology of cardiogenic components from the observed signal as the fourth derivative of a Gaussian function:
\begin{equation}
\psi_c(t)=\frac{d^4}{d t^4}\left(e^{-t^2 / 2}\right)=\left(t^4-6 t^2+3\right) e^{-t^2 / 2},
\end{equation}
with the morphology shown as the green wavelet in Fig.~\ref{fig:match_filter}.

After applying the matched filter in (\ref{equ:match}) and template $\psi_c$, the filter response $y_f(t)$ is obtained as shown in Fig.~\ref{fig:match_filter} with the prominent peaks (red dots) representing the candidates for the high-quality BSG cycles. However, in real-life environments, the morphology of BSG cycles can vary substantially across subjects, body postures and mattress conditions. As a result, the actual waveform may deviate from the predefined template shape in terms of relative peak amplitude, temporal width and local asymmetry, leading to missed or false detection of valid cycles. In addition, the cardiogenic template mainly emphasizes a generic curvature pattern and cannot fully capture subject-specific or recording-specific morphological characteristics embedded in the observed BSG signals. Therefore, using only a fixed analytical template may limit the adaptability and discriminative power of the filter.

\subsubsection{Matched Filter with Dynamic Template Learned from the Current Segment}
To further quantify the reliable cardiogenic components contained in the current segment, another matched filter is adopted based on a dynamic template $\psi_d$ learned from the current candidate cycles. Firstly, all the candidates are extracted and denoted by their peak locations $\{t_i\}_{i=1}^{M}$, as shown by the grey signals in Fig.~\ref{fig:match_filter}. Each candidate cycle can be expressed as:
\begin{equation}
p_i(t) = s(t_i + t),\ \text{with} \ t \in [-T_w/2,\; T_w/2]
\end{equation}
where $T_w$ denotes the predefined window length. Considering that these segments are selected based on the previous matched filter, they are expected to contain relatively clean cardiogenic components and therefore serve as suitable candidates for template learning.

To construct the dynamic template, a straightforward approach is to directly average the extracted cycles, but such point-wise averaging may blur the representative morphology and weaken the sharpness of physiologically meaningful waveform components due to temporal mismatch, as shown by the orange signal in Fig.~\ref{fig:match_filter}. Therefore, we use linear time warping (LTW) to rescale each candidate prior to template aggregation.

Specifically, let $L_i$ denote the target length assigned to the $i$-th candidate cycle after warping, and the LTW-transformed $p_i$ can be expressed as:
\begin{equation}
\tilde{p}_i^{(L_i)}(\tau) = p_i\!\left(\frac{T_w}{L_i}\tau\right),\ \text{with}\ \tau \in \left[-L_i/2,\, L_i/2\right],
\end{equation}
where the ratio ${L_i}/{T_w}$ characterizes the temporal scaling factor of the $i$-th cycle. In practice, the warped signal is obtained by linear interpolation, allowing each candidate to be stretched or compressed along the time axis while preserving its overall waveform morphology.

To obtain a common domain for template learning, each warped cycle $\tilde{p}_i^{(L_i)}$ is truncated or padded to the original window length $T_w$ over $\tau\in[-T_w/2,T_w/2]$. We denote the resulting fixed-length cycle by $\hat{p}_i^{(L_i)}$. The dynamic template learning problem is then formulated to find the optimal scaling lengths as
\begin{equation}
\{L_i^\ast\}_{i=1}^{M}=\arg\min_{\{L_i\}_{i=1}^{M}}\sum_{i=1}^{M}\left\|\hat{p}_i^{(L_i)}-\psi_d\right\|_2^2 ,
\label{eq:ltw_template_obj}
\end{equation}
with
\begin{equation}
\psi_d=\frac{1}{M}\sum_{i=1}^{M}\hat{p}_i^{(L_i)}.
\end{equation}

The objective in (\ref{eq:ltw_template_obj}) seeks a set of scaling lengths $\{L^*_i\}_{i=1}^{M}$ together with a shared template $\psi_d$, such that all candidate cycles become as close as possible to the same representative morphology. In addition, the scaling ratio ${L_i}/{T_w}$ is restricted within $\pm15\%$ to prevent a trivial solution. In this way, the learned template is less affected by local temporal mismatch across cycles, while the resulting template ${\psi}_d(t)$ can be interpreted as a data-adaptive representative heartbeat morphology learned from the current segment, as shown by the red signal in Fig.~\ref{fig:match_filter}.

\subsubsection{Assessment of Signal Quality with Feature Enhancement}\label{sec:quality_assessment}
After the quality evaluation by two matched filters using both cardiogenic and dynamic templates, all the high-quality BSG cycles are identified as shown in Fig.~\ref{fig:match_filter}, and the quality control can be performed by counting the recognized cardiogenic cycles in the current segment. The matched-cycle count indicates the repeated cardiogenic information available for extracting a shared cardiac pattern and estimating BP. An absolute count threshold is used because a lower heart rate does not necessarily reduce the number of recognizable cycles needed by the model. In addition, the Hilbert envelope is extracted for the preserved segments to enhance the cardiac feature, as used by many previous studies~\cite{wang2024accurate,zhang2023overview}.

\subsection{Physical Model for 3D Wave Propagation}
To further model the intrinsic relationships between high-quality triaxial BSG signals and cardiogenic excitation, the body-bed system can be modeled as an equivalent viscoelastic medium under small-amplitude mechanical excitation (i.e., cardiogenic forces $\mathbf{f_c}$)~\cite{Carcione2022}. The wave propagation theory provides adequate tools to capture wave attenuation, dispersion, and phase delay~\cite{luporini2020architecture}, with the dominant dynamics mainly governed by elastic and viscous properties. This section will construct the constitutive model for cardiogenic excitation and the body-bed system as a theoretical foundation of the physics-constrained deep learning layer. However, only the key steps are presented in the main text for clarity, while the detailed derivation is provided in the supplemental material.

\subsubsection{Wave Propagation Model for the Body-bed System}
By modeling the body-bed system as an equivalent viscoelastic medium, the propagation of cardiogenic vibration can be expressed as
\begin{equation}\label{equ:combine1}
\rho \frac{\partial^2 \mathbf{u}}{\partial t^2} = \nabla \cdot (\boldsymbol{E} : \nabla_s \mathbf{u}) + \nabla \cdot (\boldsymbol{\eta} : \nabla_s \dot{\mathbf{u}}) + \mathbf{f}_c,
\end{equation}
where $\mathbf{u}(\xi,t)\in\mathbb{R}^3$ is the displacement field, $\rho$ is the density, $\boldsymbol{E}$ and $\boldsymbol{\eta}$ denote the elastic and viscous parameters, and $\mathbf{f}_c$ represents the cardiogenic excitation force. This equation shows that wave propagation in the body-bed system is jointly governed by inertial, elastic, and viscous effects.

\subsubsection{Finite-Dimensional Dynamic System}
Using the finite element method (FEM), the wave propagation model in (\ref{equ:combine1}) can be transformed into a finite-dimensional second-order dynamical system:
\begin{equation}\label{equ:govern_ode1}
\mathbf{M}\ddot{\mathbf{q}}(t)+\mathbf{C}\dot{\mathbf{q}}(t)+\mathbf{K}\mathbf{q}(t)=\mathbf{F}_c(t),
\end{equation}
where $\mathbf{q}(t)$ denotes the nodal displacement states, $\mathbf{M}$, $\mathbf{C}$, and $\mathbf{K}$ are the mass, damping, and stiffness matrices, respectively, and $\mathbf{F}_c(t)$ is the equivalent force induced by cardiogenic excitation. The detailed expressions can be found in the supplemental material.

\subsubsection{Model Reduction for the Body-bed System}\label{sec:ss_model}
According to the previous FEM discretization, the original PDE in (\ref{equ:combine1}) is simplified as the ordinary differential equation (ODE) model in (\ref{equ:govern_ode1}). However, since the body-bed system is assembled from a large number of nodes, the resulting state dimension is too high for efficient learning and inference. Therefore, modal truncation is adopted to derive a reduced-order model.

Specifically, we assume that the measured BSG signals are mainly dominated by a limited number of low-frequency vibration modes. After removing unobservable rigid-body modes, the system matrices are assumed to admit a valid modal decomposition. In addition, proportional (Rayleigh) damping is adopted with the damping matrix expressed as a linear combination of the mass and stiffness matrices as:
\begin{equation}
\mathbf{C}=\alpha \mathbf{M}+\beta \mathbf{K},
\end{equation}
where $\alpha$ and $\beta$ are the Rayleigh damping coefficients.

Based on the generalized eigenvalue problem of the discretized system, the nodal displacement is approximated by the first $r$ dominant modes as:
\begin{equation}
\mathbf{q}(t)\approx \mathbf{\Phi}_r \mathbf{z}(t),
\end{equation}
where $\mathbf{\Phi}_r$ contains the retained mode shapes and $\mathbf{z}(t)\in\mathbb{R}^r$ denotes the reduced modal coordinates. By projecting the full-order dynamics onto this modal subspace, the original high-dimensional system is reduced to:
\begin{equation}
\ddot{\mathbf{z}}(t)+\mathbf{D}\dot{\mathbf{z}}(t)+\mathbf{\Omega}^2\mathbf{z}(t)=\mathbf{F}_r(t),
\end{equation}
with
\begin{equation}\label{equ:rom_terms1}
\begin{aligned}
\mathbf{F_r}(t)&=\mathbf{\Phi}_r^{\top}\mathbf{F}_c(t), \\
\mathbf{\Omega}^{2}&=\mathrm{diag}(\omega_1^2,\dots,\omega_r^2),\\
\mathbf{D}&=\alpha\mathbf{I}+\beta\mathbf{\Omega}^{2},
\end{aligned}
\end{equation}
where $\mathbf{\Omega}^2$ is the diagonal matrix of retained modal frequencies, $\mathbf{D}$ is the reduced damping matrix, and $\mathbf{F}_r(t)$ is the reduced generalized force.

Finally, by defining the state vector as $\mathbf{x}(t)=[\mathbf{z}(t)^\top,\dot{\mathbf{z}}(t)^\top]^\top$, the reduced-order system can be written in state-space form as:
\begin{equation}\label{equ:ss1}
\dot{\mathbf{x}}(t)=
\begin{bmatrix}
\mathbf{0} & \mathbf{I}\\
-\mathbf{\Omega}^{2} & -\mathbf{D}
\end{bmatrix}\mathbf{x}(t)
+
\begin{bmatrix}
\mathbf{0}\\
\mathbf{I}
\end{bmatrix}\mathbf{F}_r(t),
\end{equation}
and the measured triaxial BSG signal is expressed as:
\begin{equation}\label{equ:ss2}
\mathbf{y}(t)\approx
\begin{bmatrix}
\mathbf{0} & \mathbf{S}\mathbf{\Phi}_r
\end{bmatrix}\mathbf{x}(t),
\end{equation}
where $\mathbf{S}$ denotes the observation matrix mapping the latent structural response to the sensor channels. In practice, only the first few dominant modes within the effective BSG frequency band are retained, and the reduced model in (\ref{equ:ss1}) preserves the principal low-frequency dynamics while remaining computationally tractable for the later design of the physics-constrained layer.
\begin{figure}[!t]
    \centering
    \captionsetup[subfloat]{farskip=0pt,nearskip=0pt}
    \subfloat{\label{fig:physical_fitting}%
        \begin{minipage}{\columnwidth}\vspace*{9pt}
        \begin{overpic}[width=\columnwidth]{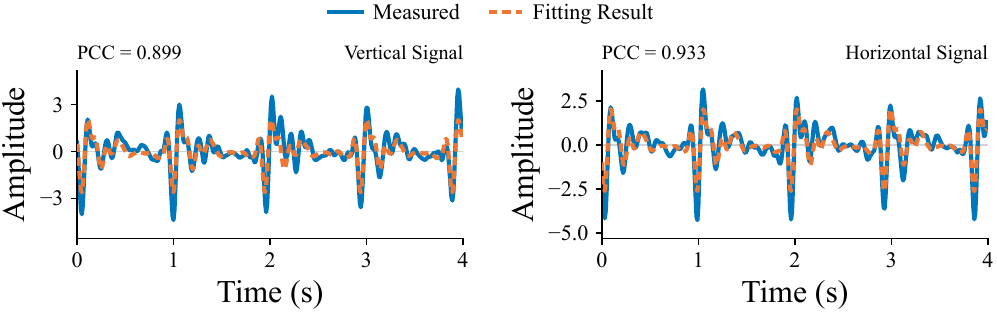}
            \put(0,34.32){\makebox(0,0)[tl]{{\footnotesize (a)}}}
        \end{overpic}\end{minipage}}\\[4pt]
    \subfloat{\label{fig:physical_fit_pcc}%
        \begin{minipage}{\columnwidth}\vspace*{9pt}
        \begin{overpic}[width=\columnwidth]{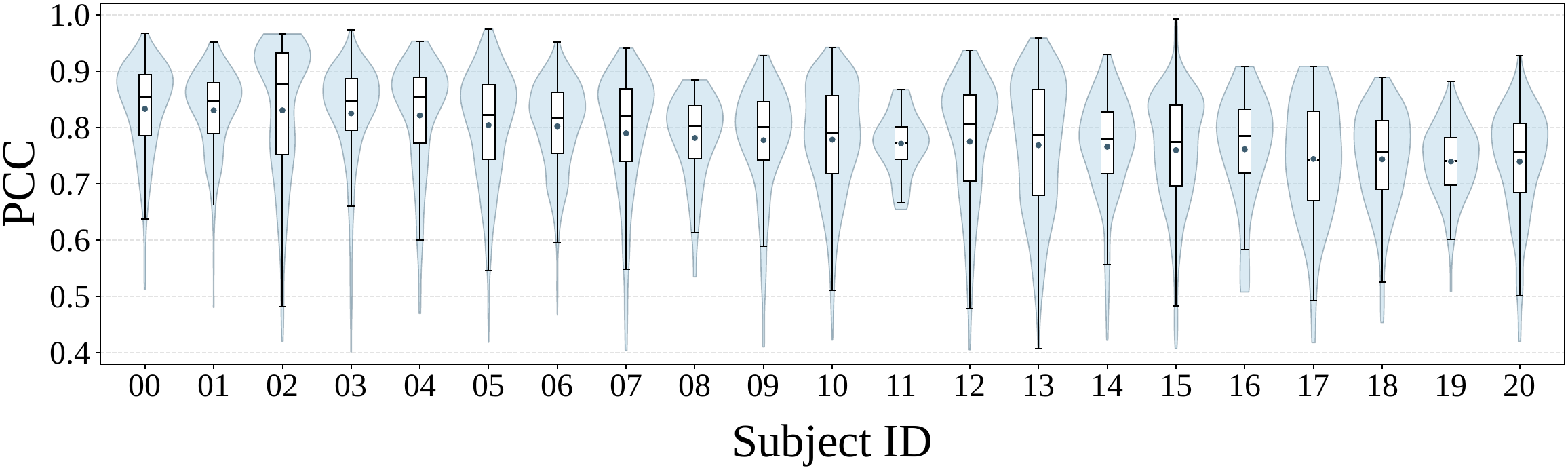}
            \put(0,33.03){\makebox(0,0)[tl]{{\footnotesize (b)}}}
        \end{overpic}\end{minipage}}\\[4pt]
    \subfloat{\label{fig:modal_frequency_distributions}%
        \begin{minipage}{\columnwidth}
        \noindent\makebox[\linewidth][l]{{\footnotesize (c)}}\par\vspace{2pt}
        \centering
        \includegraphics[width=0.9\columnwidth]{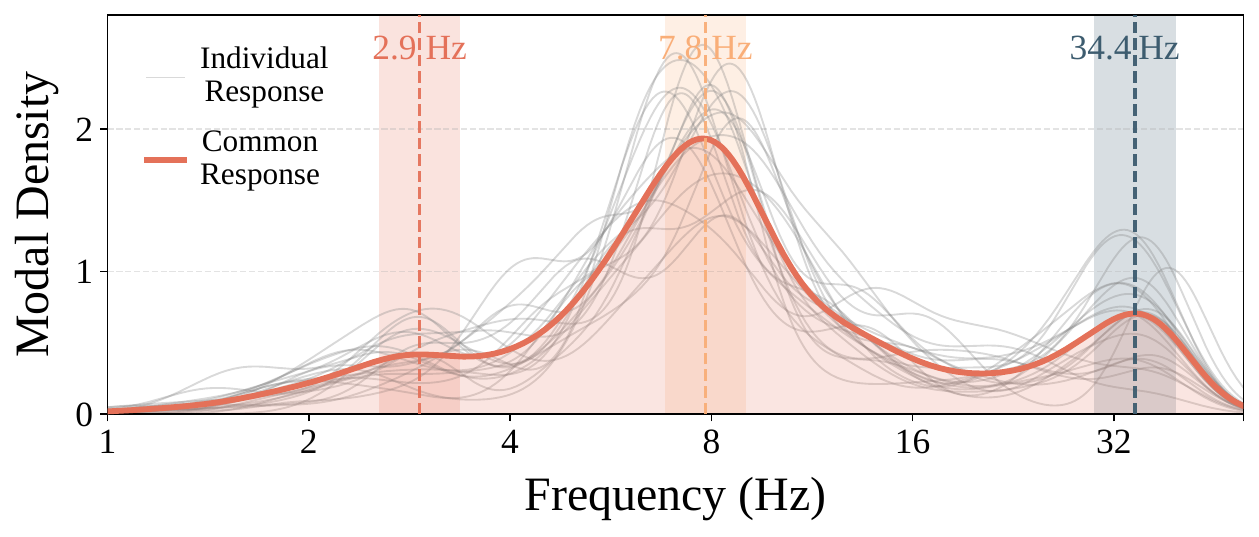}
        \end{minipage}}
    \caption{Signal-level assessment of the reduced model: (a) Measured signals and fitting results; (b) Overall fitting Pearson correlation coefficient (PCC); (c) Modal frequency distributions.}
    \label{fig:physical_validation}
\end{figure}
\subsubsection{Rationale and Signal-Level Assessment of the Reduced Model}\label{sec:reduced_model_validation}
Although tissue and mattress responses can be nonlinear and time-varying, structured latent dynamics can support prediction of complex temporal behavior without an explicitly specified physical system~\cite{cheng2026information,xiong2026kuma}. The reduced model in (\ref{equ:ss1}) retains inertia, damping, and stiffness, providing a physically motivated inductive bias. To assess (\ref{equ:ss1}), we fitted the reduced model to measured BSG segments using a leave-one-subject-out (LOSO) procedure. The fitted waveforms follow the main oscillatory patterns of the measurements (Fig.~\ref{fig:physical_validation}(a)), with fitting PCC varying across subjects (Fig.~\ref{fig:physical_validation}(b)), and recurring peaks appear near $2.9$, $7.8$, and $34.4$~Hz (Fig.~\ref{fig:physical_validation}(c)). These results support the compact dynamical structure under the studied bed configuration. In Section~\ref{sec:phy_DL}, we regularize latent states by using the second-order form of (\ref{equ:ss1}) without adopting the numerical modal parameters of the reduced model.

\subsection{Physics-constrained Deep Learning Model}\label{sec:phy_DL}
To provide a clear reference for subsequent model comparison, the plain encoder-decoder architecture is adopted as the benchmark backbone in this study, as shown in Fig.~\ref{fig:overview}(c), while the state-space models in (\ref{equ:ss1}) are embedded as physics-constrained layers later in this section.
\subsubsection{Benchmark Architecture Design}
Table~\ref{tab:encoder_structure} illustrates the encoder design. A 1-D ResNet-18 architecture with residual blocks is adopted to extract cardiac information~\cite{he2016identity}. Although many advanced encoders have been proposed for time-series processing, ResNet remains widely used for real-world signal analysis, including machinery fault diagnosis~\cite{chen2026cows}, radar-based cardiac signal processing~\cite{zhang2025high}, and ECG analysis~\cite{boubekki2026explaining}. The shared local kernels make ResNet well suited to capturing recurring waveform patterns in quasi-periodic BSG signals. Its relatively simple convolutional architecture also supports efficient inference, making it a practical choice for future real-time processing.

Three identically configured encoders process the BSG axes separately. Each encoder receives the waveform and its Hilbert envelope as two channels, forming an input of size $(N,2,1000)$, where $N$ is the batch size. Convolution, max pooling, and residual blocks produce features of size $(N,128,32)$ for cross-axis alignment and fusion in the physics-constrained layer (Section~\ref{sec:phy_layer_design}). The decoder then processes the fused features using convolutional (Conv) and transposed convolutional (TransConv) blocks. The resulting feature map is flattened and passed through Linear blocks, followed by a final Linear layer predicting SBP, DBP, and MAP. Each Conv, TransConv, or Linear block includes the corresponding layer, batch normalization (BN), and a rectified linear unit (ReLU).

\begin{table}[t]
\caption{Structure and Parameters for the Encoder and Decoder}
\label{tab:encoder_structure}
\centering
\begin{tabular}{l c c}
\toprule
\multicolumn{1}{c}{\multirow{2}*{\textbf{\makecell[c]{Encoder \\ Layers}}}} & \textbf{Parameters}  & \textbf{Output Shape} \\
& ($C_{in}$, $C_{out}$, $K$, $S$)$^1$ & $N$: Batch Size \\
\toprule
Input signal & -- & $(N, 2, 1000)$ \\
\midrule
\textbf{Encoder} & & \\
\ \ Conv1d & $(2, 64, 7, 2)$ & $(N, 64, 500)$ \\
\ \ MaxPool & $(64, 64, 3, 2)$ & $(N, 64, 250)$ \\
\ \ Residual Block $\times 2$ & $(64, 16, 3, 1)$ & $(N, 16, 250)$ \\
\ \ Residual Block $\times 2$ & $(16, 32, 3, 2)$ & $(N, 32, 125)$ \\
\ \ Residual Block $\times 2$ & $(32, 64, 3, 2)$ & $(N, 64, 63)$ \\
\ \ Residual Block $\times 2$ & $(64, 128, 3, 2)$ & $(N, 128, 32)$ \\
\toprule
\textbf{Decoder} & & \\
\ \ Conv Block & $(128, 32, 3, 1)$ & $(N, 32, 32)$ \\
\ \ Conv Block & $(32, 16, 3, 1)$ & $(N, 16, 32)$ \\
\ \ Conv Block & $(16, 8, 3, 1)$ & $(N, 8, 32)$ \\
\ \ Conv Block & $(8, 4, 3, 1)$ & $(N, 4, 32)$ \\
\ \ TransConv Block & $(4, 4, 3, 2)$ & $(N, 4, 62)$ \\
\ \ TransConv Block & $(4, 8, 3, 2)$ & $(N, 8, 122)$ \\
\ \ TransConv Block & $(8, 8, 3, 1)$ & $(N, 8, 120)$ \\
\ \ Linear Block & $(120\times 8, 1024, -, -)$ & $(N, 1024)$ \\
\ \ Linear Block & $(1024, 512, -, -)$ & $(N, 512)$ \\
\ \ Linear Layer & $(512, 3, -, -)$ & $(N, 3)$ \\
\midrule
Output & $-$ & $(N, 3)$ \\
\bottomrule
\multicolumn{3}{l}{$^1C_{in}$: input channels; $C_{out}$: output channels; $K$: kernel size; $S$: stride.} \\
\end{tabular}
\end{table}

\begin{figure*}[tb] 
    \centering 
    \includegraphics[width=1.6\columnwidth]{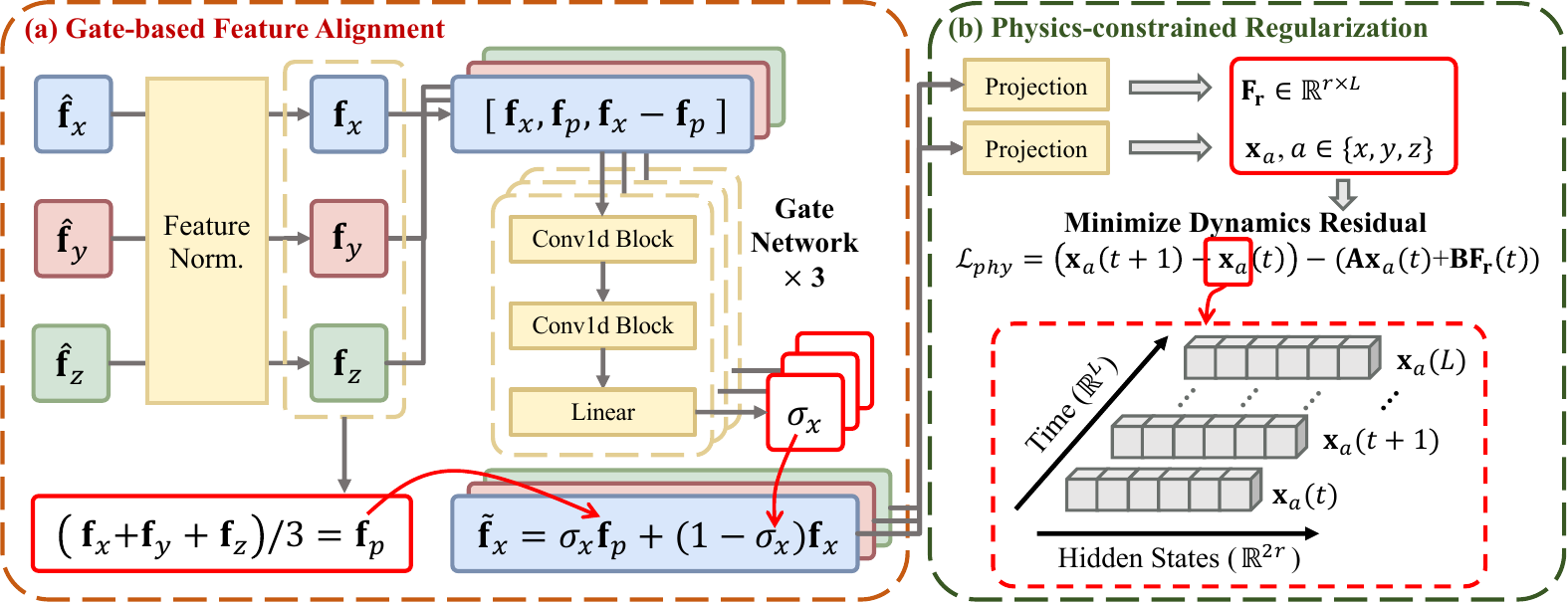}
    \caption{Design of physics-constrained layer: (a) Gate-based feature alignment to aggregate triaxial features; (b) Physics-constrained regularization to govern the feature evolution during network training.}
    \label{fig:phy_layer} 
\end{figure*}

\subsubsection{Physics-constrained Layer Design}\label{sec:phy_layer_design}
To explicitly embed the reduced-order physical model derived in Section~\ref{sec:ss_model} into the deep learning pipeline, a physics-constrained layer is introduced between the encoder and decoder to align the latent representations extracted from the three BSG axes, as shown in Fig.~\ref{fig:phy_layer}. The inputs of this layer are the latent features produced by the three encoders $\mathbf{\hat{f}}_x,\mathbf{\hat{f}}_y,\mathbf{\hat{f}}_z \in \mathbb{R}^{C\times L}$, as shown in Table~\ref{tab:encoder_structure}, while these features will be aligned through two steps to preserve axis-specific informative components and suppress distortion-induced inconsistency.

\textbf{Gate-based Feature Alignment:}
The latent features extracted from the X-, Y-, and Z-axis are firstly normalized along the channel dimension to reduce scale discrepancy, yielding ${\mathbf{f}}_x,{\mathbf{f}}_y,{\mathbf{f}}_z$ and the shared prototype $\mathbf{f}_p$ as shown in Fig.~\ref{fig:phy_layer}(a). Then, for each axis $a\in\{x,y,z\}$, a time-varying gate $\sigma_a\in[0,1]$ is learned from the concatenated feature tuple $[{\mathbf{f}}_a,\mathbf{f}_p,{\mathbf{f}}_a-\mathbf{f}_p]$ through a gate network, and the aligned feature is formulated as:
\begin{equation}
\tilde{\mathbf{f}}_a(t)=\sigma_a\mathbf{f}_p(t)+\left(1-\sigma_a\right){\mathbf{f}}_a(t).
\end{equation}
Given that cardiogenic vibration energy may redistribute across axes and yield different yet still informative observations in triaxial BSG measurements, the gate-based soft alignment allows the model to selectively preserve the informative axis-specific components, while pulling the feature toward the shared prototype when distortion or energy leakage weakens its cardiogenic consistency.

\textbf{Physics-constrained Latent State Regularization:}
After gate-based feature alignment, the aligned features across the three axes are further regularized in a low-dimensional latent state to enhance their physical consistency during deep learning model training. Specifically, the aligned feature $\tilde{\mathbf{f}}_a\in\mathbb{R}^{C\times L}$ is projected into a latent state sequence $\mathbf{x}_a(t)\in\mathbb{R}^{2r}$ through an individual projection network, where $r$ denotes the reduced modal dimension. Meanwhile, the fused feature $\mathbf{f}_{\mathrm{fused}}=(\tilde{\mathbf{f}}_x+\tilde{\mathbf{f}}_y+\tilde{\mathbf{f}}_z)/3$ is projected to a shared latent excitation $\mathbf{F}_r\in\mathbb{R}^{r\times L}$, representing the common cardiogenic driving effect underlying the three-axis measurements.

The encoder has a total temporal stride of $32$, giving a latent sampling interval of $\Delta t=32/100=0.32$~s. The implementation uses normalized time $\tau=t/\Delta t$ with a unit step $\Delta\tau=1$. Based on the reduced-order dynamics in (\ref{equ:ss1}), the latent states from all three axes $a\in\{x,y,z\}$ are constrained using the forward Euler approximation:
{
\begin{equation}
(\mathbf{x}_{a,k+1}-\mathbf{x}_{a,k})/\Delta\tau\approx \mathbf{A}\mathbf{x}_{a,k}+\mathbf{B}\mathbf{F}_{r,k},
\end{equation}}%
where $k$ indexes latent samples. All quantities in (17) use normalized time, with the latent state convention $\mathbf{x}_a=[\mathbf{z}_a^{\top},(d\mathbf{z}_a/d\tau)^{\top}]^{\top}$. The matrices $\mathbf{A}$ and $\mathbf{B}$ retain the block structure in (\ref{equ:ss1}), with $\mathbf{\Omega}^{2}=(\Delta t)^2\mathbf{\Omega}_{\mathrm{s}}^{2}$, $\mathbf{D}=\Delta t\,\mathbf{D}_{\mathrm{s}}$, and $\mathbf{F}_r=(\Delta t)^2\mathbf{F}_{r,\mathrm{s}}$. The subscript $\mathrm{s}$ denotes the corresponding quantities in physical time measured in seconds. Therefore, the learned features are not forced to share identical latent states, but are encouraged to follow the same underlying dynamical law by minimizing the dynamics residual $\mathcal{L}_{phy}$, as shown in Fig.~\ref{fig:phy_layer}(b).

In implementation, the proposed layer contains several groups of learnable parameters. The state and excitation projection networks each use two Conv Blocks with $2r$ and $r$ output channels, respectively. The normalized-time modal angular frequencies $\{\omega_1,\dots,\omega_r\}$ and two nonnegative Rayleigh damping scalars $\alpha$ and $\beta$ determine $\mathbf{A}$ through the structure in (\ref{equ:rom_terms1}) and (\ref{equ:ss1}). The angular frequencies are expressed in radians per latent time step, corresponding to $f_i=\omega_i/(2\pi\Delta t)$ in Hz as a parameter-unit conversion. This conversion does not identify the learned latent frequencies with the signal-level peaks in Fig.~\ref{fig:physical_validation}(c).

\textbf{Overall Loss Function:}
Finally, the overall training loss is defined as $\mathcal{L}_{\mathrm{all}} = \mathcal{L}_{{BP}} + \lambda \mathcal{L}_{{phy}}$, where $\mathcal{L}_{\mathrm{BP}}$ denotes the regression loss between the predicted and ground-truth BP values, $\mathcal{L}_{\mathrm{phy}}$ is the physics-guided regularization term, and $\lambda$ is a weighting coefficient that balances the two loss components. To avoid degenerate latent representations, we adopt the structural consistency loss and entropy regularization in~\cite{cheng2026information} to measure discrepancies between computed and predicted latent-state increments while preserving latent diversity.

\section{Details of Experiments and Dataset}\label{sec:dataset}
\subsection{Dataset Collection and Description}
\begin{figure*}[tb] 
    \centering 
    \subfloat[]{\label{fig:bed_illustration}\includegraphics[width=0.98\columnwidth]{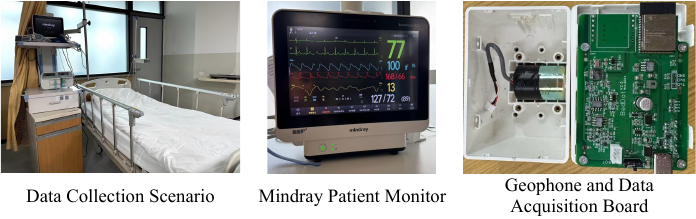}}\hspace*{0.02\columnwidth}
    \subfloat[]{\label{fig:MAP_kde}\includegraphics[width=0.52\columnwidth]{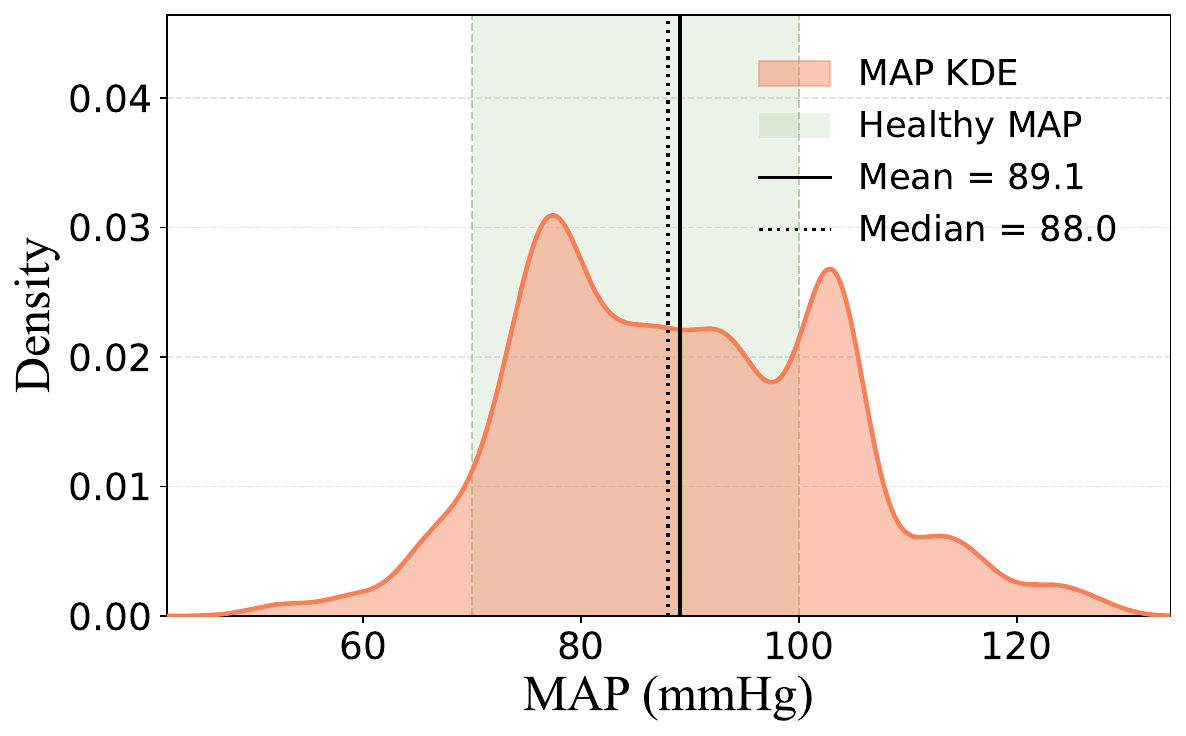}}
    \subfloat[]{\label{fig:MAP_age}\includegraphics[width=0.52\columnwidth]{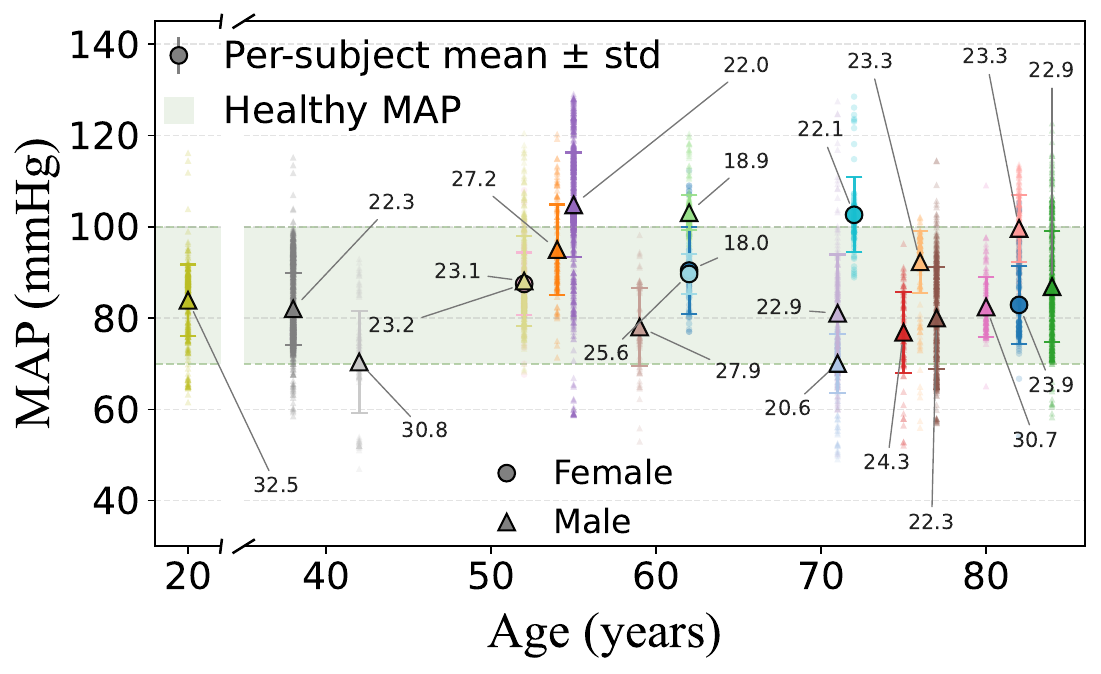}}
    \caption{Overview of the dataset: (a) Data collection overview; (b) MAP Distribution; (c) MAP by Age, with the numerical annotations indicating BMI.}
    \label{fig:distribution} 
\end{figure*}

Fig.~\subref*{fig:bed_illustration} illustrates the data collection scenarios in a hospital setting, and three seismic sensors are mounted under the bed frame at different orientations for 3-axis sensing. The same RacoTech geophone~\cite{rgi45hz} and data acquisition board are used to form a sensing system with a lower cutoff frequency of approximately $0.7$~Hz, and the adopted geophone consists of a fixed magnetic core enclosed by wire coils as a transducer for subtle vibrations. This geophone-based sensor system has been validated for faithful cardiac or sleep monitoring in our previous studies~\cite{chen2025selfdenoiser,li2020smart,song2026sleepapnea}, with no noticeable synchronization problems observed in these applications.

The dataset contains a total of $162$ hours of synchronous data collected from $21$ people aged $20$ to $84$ ($16$ male and $5$ female participants). The medical-grade Mindray BeneVision N12 provides continuous monitoring of ABP, ECG, and PPG signals. These signals are uniformly resampled to $100$~Hz for subsequent analysis. Abnormal ABP measurements are handled by the algorithms embedded in the monitor without additional artifact rejection. The monitor also automatically calculates and outputs SBP, DBP, and MAP values once per second. These values are averaged over each corresponding $10$-sec BSG segment to form the reference labels. In addition, CO and SV data are available from PiCCO~\cite{litton2012picco} to characterize individual hemodynamic conditions, such as TPR.

The overall data distribution is summarized in Fig.~\ref{fig:distribution}(b), while Fig.~\ref{fig:distribution}(c) presents MAP by age with sex and body mass index (BMI) information. Although most subjects are within a relatively healthy BP range, the naturally collected hospital dataset still covers a wider MAP distribution than many controlled laboratory datasets. Such a broad BP range is important for evaluating whether a model truly learns BP-related cardiogenic features, rather than obtaining seemingly good performance by predicting values close to the population mean. In addition, the dataset covers a wide age range and diverse BP conditions, enabling a more comprehensive evaluation of model robustness across different physiological states.

\subsection{Implementation Details}
\subsubsection{Model Training Strategy}
The deep learning model is coded in PyTorch and trained on NVIDIA RTX 4090 (24GB) for $100$ epochs with batch size $512$. The recordings are divided into non-overlapping $10$-sec segments. A LOSO protocol is used for model evaluation, with $1$ subject held out for testing and the remaining $20$ subjects alternately selected for training or validation (i.e., $20$-fold cross-validation). A new model is trained for each LOSO split. Candidate values of the physics-loss weight $\lambda$ and the matched-cycle threshold are compared in ablation experiments using validation results from the $20$ non-test subjects, and the selected settings are fixed before the held-out subject is evaluated. All experiments are repeated $5$ times on the same $21$ subjects, and mean results are reported, with $95\%$ confidence interval half-widths shown for the Overall scores in Table~\ref{tab:bp_results}.

\subsubsection{Baseline and Benchmark Evaluation}
\label{sec:baseline_benchmark}
Hyde~\cite{huang2026hyde}, a recently published single-axis BCG model based on multi-task domain adaptation, serves as the reference baseline for calculating overall relative improvement in Table~\ref{tab:bp_results}. The comparison further includes the PPG/ECG-based models PITN~\cite{wang2025pitn} and MSTNN~\cite{fan2023fewshotbp}, with PITN employing physics-informed temporal learning and MSTNN employing multi-modal spectro-temporal feature extraction. The single-axis BCG models FSNet~\cite{zhang2026bcg} and AI-BCG~\cite{huang2024ai} employ few-shot meta-learning and waveform and fiducial features with personalization, respectively. All methods use the same subject-level data splits, calibration procedure, and evaluation metrics.

The performance of BP prediction will be evaluated based on mean error (ME), mean absolute error (MAE), Pearson correlation coefficient (PCC) and standard deviation (STD). In addition, to provide a comprehensive assessment across multiple blood pressure prediction tasks, the averaged relative improvement over the baseline $b$ is calculated across tasks and metrics as:
\begin{equation}
\Delta m\% =\frac{1}{n}\sum_{i=1}^{n}\frac{1}{n_i}\sum_{j=1}^{n_i}s_{i,j}\frac{M_{m,i,j} - M_{b,i,j}}{M_{b,i,j}}\times 100\% ,
\label{eq:overall_score}
\end{equation}
where $n=3$ denotes the number of prediction tasks (i.e., SBP, DBP, MAP prediction), and $n_i=3$ denotes the number of metrics for task $i$ (i.e., MAE, STD, PCC). $M_{m,i,j}$ represents the performance of method $m$ on task $i$ measured by metric $j$, while $M_{b,i,j}$ denotes the corresponding performance of the Hyde baseline. $s_{i, j}=1/-1$ if higher/lower values are better for the current metric (indicated by $\uparrow / \downarrow$).

\subsubsection{Calibration for Personal Hemodynamic Difference}
According to (\ref{equ:tpr}), BP is determined by CO and TPR, where CO is directly associated with BSG signal based on Starr's necropsies~\cite{starr1959studies}, but there is no evidence to imply that the personal TPR conditions can be inferred from any non-invasive measurement. Therefore, calibration is commonly recognized as a necessary step to correct the model response to fit personal hemodynamic conditions~\cite{huang2024ai,li2025practicalbp,cao2025mmwave,shi2025mmbp}. Although some studies claimed to require only a single calibration or even no calibration, these claims were made without quantitative analysis of TPR-related effects. Thus, the previous models may directly predict the mean value to minimize the ME, and the good performance relies on a stable individual hemodynamic condition, as will be shown in the later experimental results.

Our dataset includes PiCCO-derived CO measurements and supports calculation of TPR based on (\ref{equ:tpr}). The TPR analysis provides descriptive context for individual hemodynamic variation, while the effect of calibration is evaluated through comparisons across calibration intervals. The calibration schedule should be selected according to the monitoring task and individual hemodynamic stability.

All methods use the same calibration schedule and reference samples, with a target interval of approximately $8$ hours. A calibration sample consists of a $10$-sec BSG segment paired with its corresponding reference BP values. If no valid paired sample is available at the scheduled time, calibration is deferred to the first valid paired sample available thereafter. For each BP target, each method calculates an offset as the reference value minus its prediction for the selected sample and adds this offset to subsequent predictions until the next calibration update. Only historical reference measurements are used, and calibration samples are excluded from performance evaluation. For practical deployment, intermittent cuff measurements could provide reference BP values while contactless estimation supports monitoring between these measurements.

\section{Experimental Results and Evaluations}\label{sec:result}

\subsection{Overall Performance}
\begin{figure*}[t]
    \centering
    \captionsetup[subfloat]{farskip=0pt,nearskip=0pt}
    \subfloat{\label{fig:BA_phy_20_SBP}%
                \begin{overpic}[width=0.65\columnwidth]{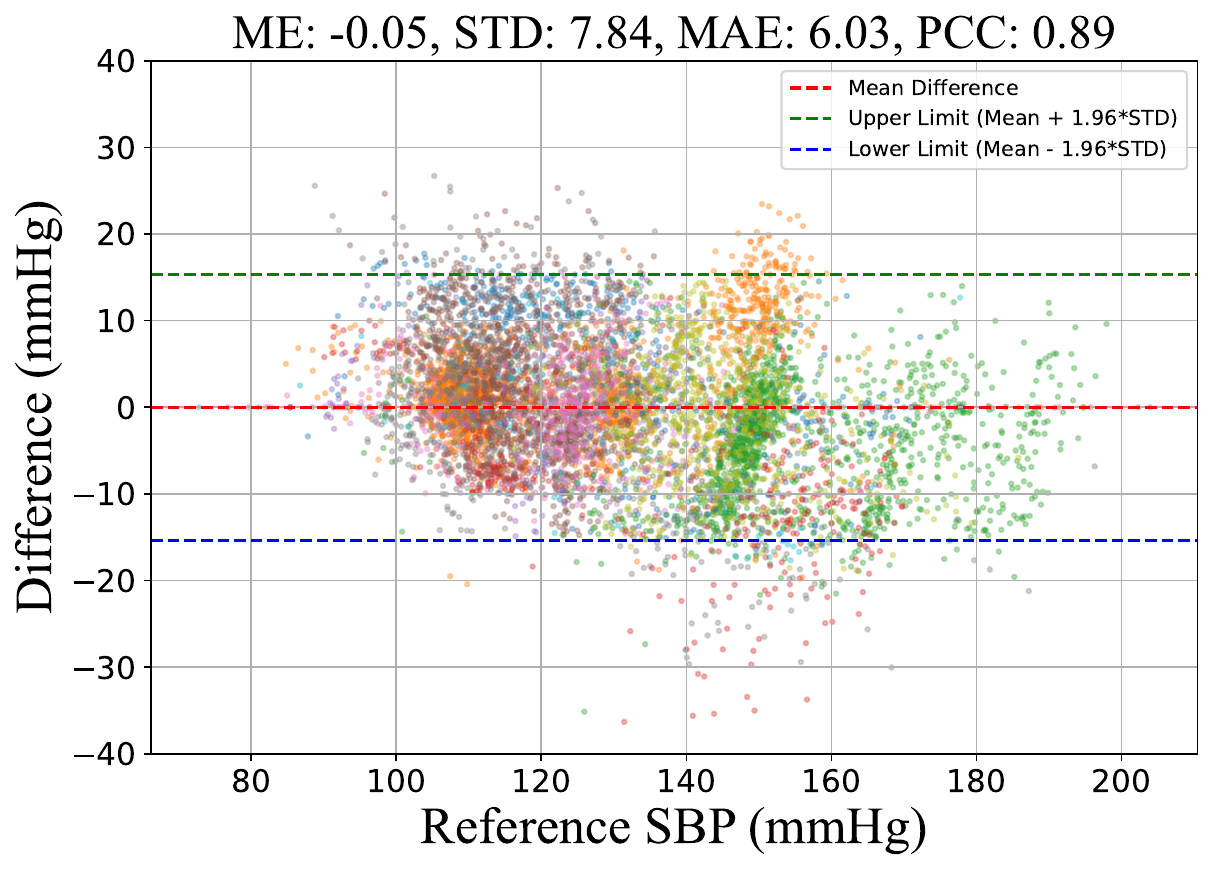}
            \put(0,71.70){\makebox(0,0)[tl]{{\footnotesize (a)}}}
        \end{overpic}} 
    \subfloat{\label{fig:BA_phy_20_DBP}%
                \begin{overpic}[width=0.65\columnwidth]{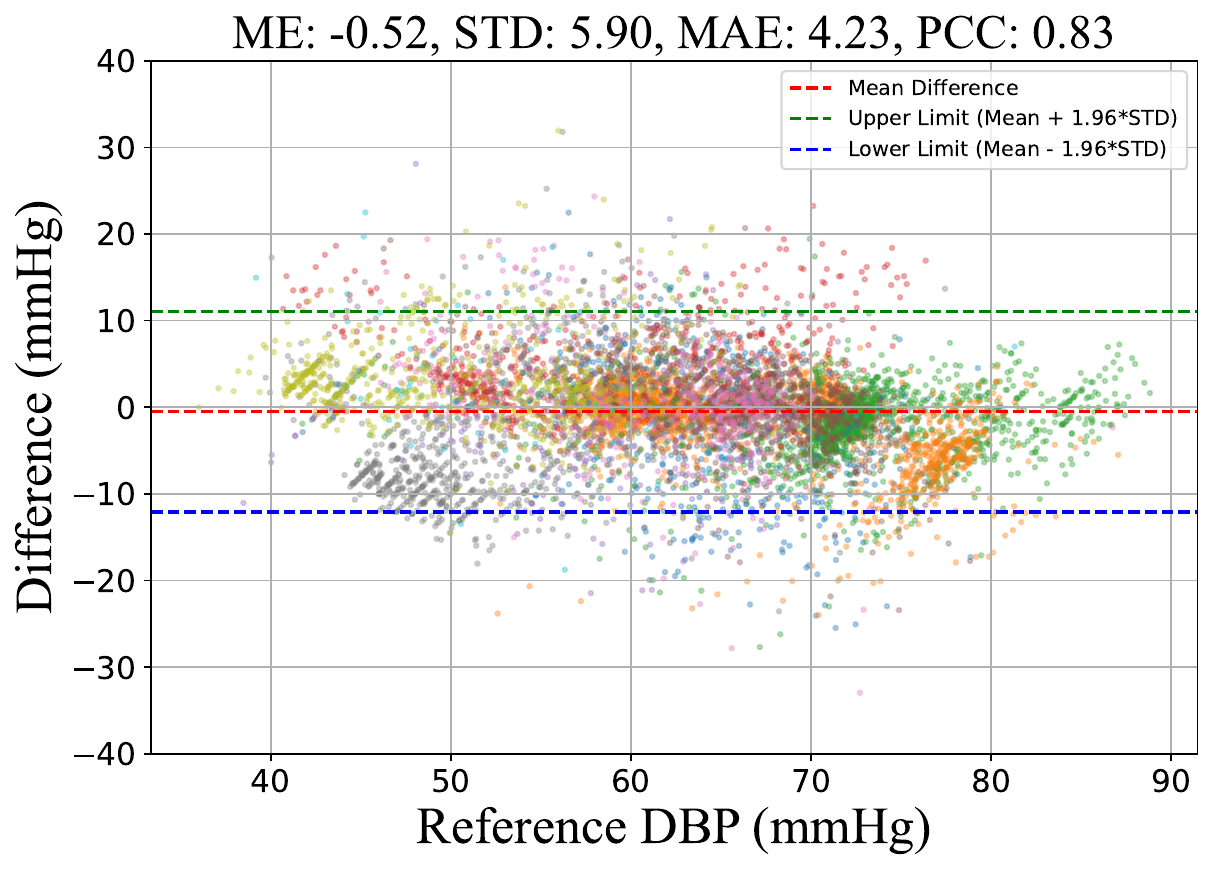}
            \put(0,71.70){\makebox(0,0)[tl]{{\footnotesize (b)}}}
        \end{overpic}} 
    \subfloat{\label{fig:BA_phy_20_MAP}%
                \begin{overpic}[width=0.65\columnwidth]{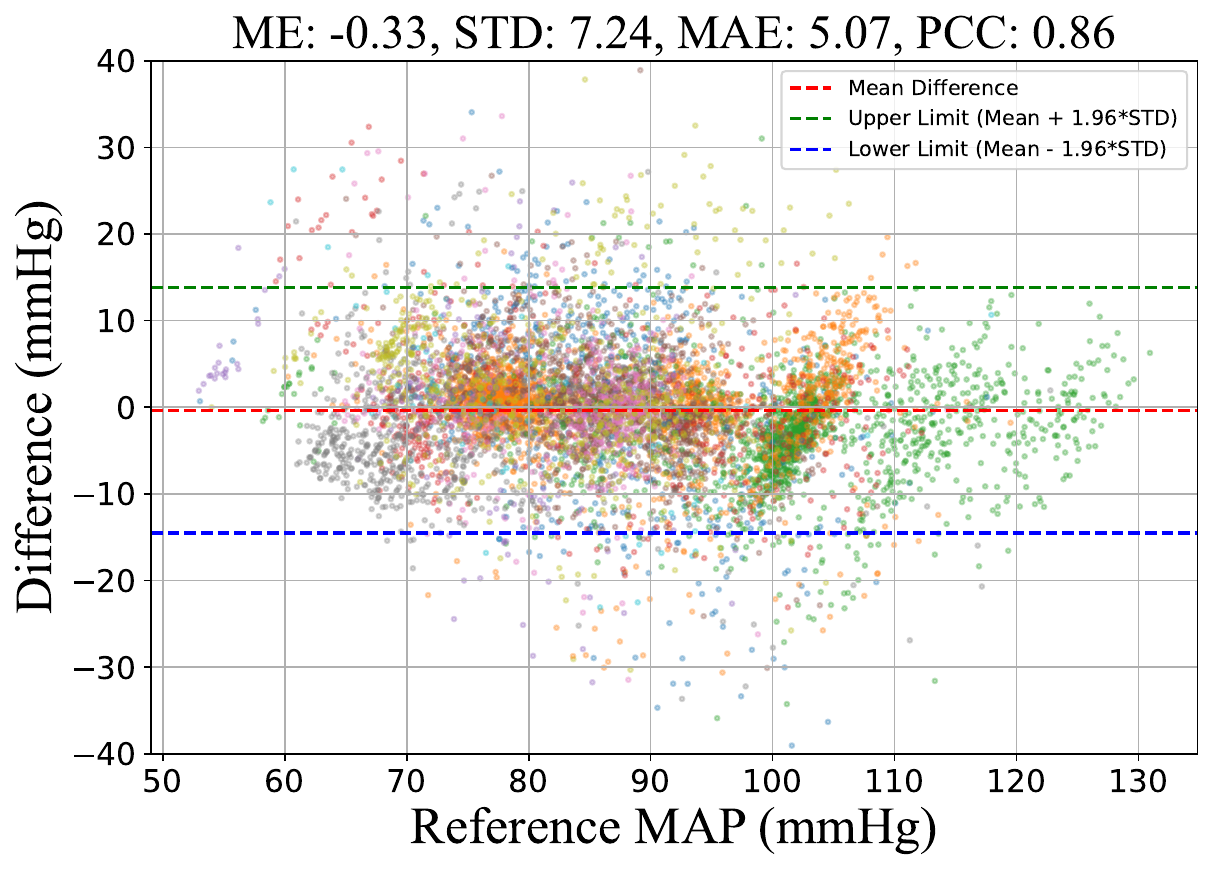}
            \put(0,71.70){\makebox(0,0)[tl]{{\footnotesize (c)}}}
        \end{overpic}} \\[3pt]
    \subfloat{\label{fig:scatter_phy_1}%
                \begin{overpic}[width=0.5\columnwidth]{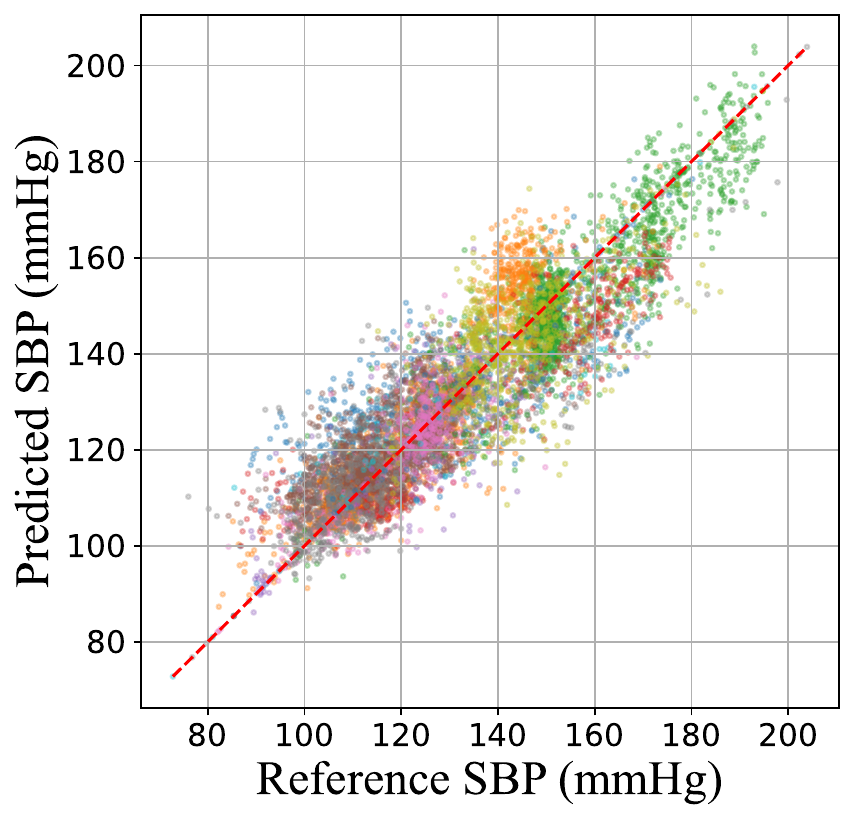}
            \put(0,96.03){\makebox(0,0)[tl]{{\footnotesize (d)}}}
        \end{overpic}} \hspace*{0.15\columnwidth}
    \subfloat{\label{fig:scatter_phy_2}%
                \begin{overpic}[width=0.5\columnwidth]{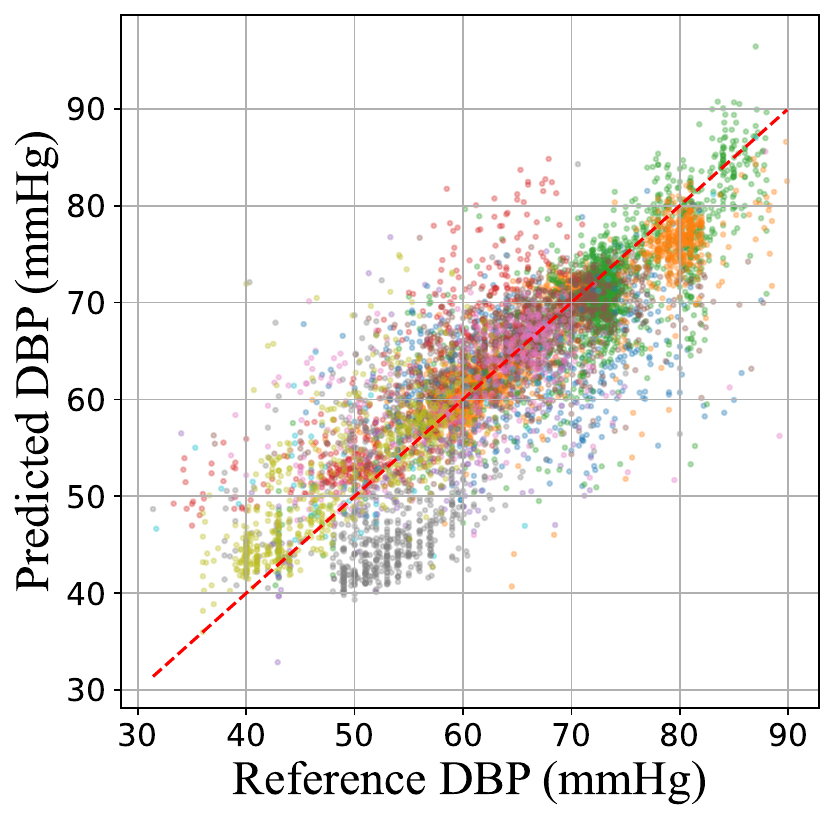}
            \put(0,98.32){\makebox(0,0)[tl]{{\footnotesize (e)}}}
        \end{overpic}} \hspace*{0.15\columnwidth}
    \subfloat{\label{fig:scatter_phy_3}%
                \begin{overpic}[width=0.5\columnwidth]{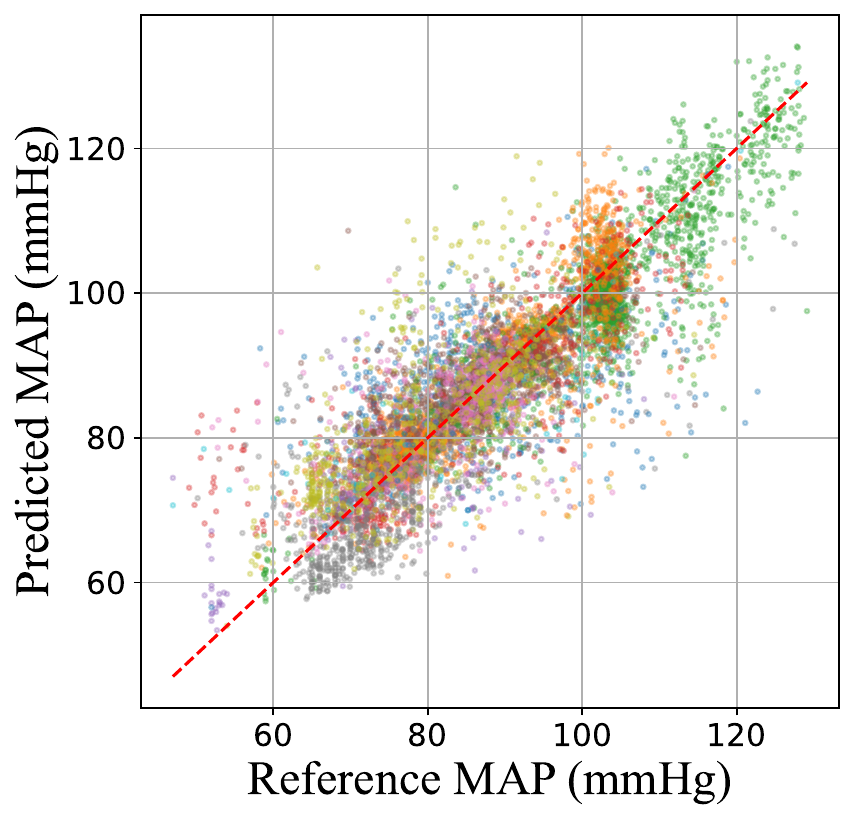}
            \put(0,96.03){\makebox(0,0)[tl]{{\footnotesize (f)}}}
        \end{overpic}}
    \caption{Performance overview of Phy-BP (Full) with different colors for different subjects: (a) - (c) Bland-Altman plot for SBP, DBP and MAP; (d) - (f) Correlation plot for SBP, DBP and MAP.}
    \label{fig:result_phy}
\end{figure*}

\begin{figure}[!t]
    \centering
    
    \captionsetup[subfloat]{farskip=0pt,nearskip=0pt}
    \subfloat{\label{fig:subject_level_mae}%
        \begin{overpic}[width=\columnwidth]{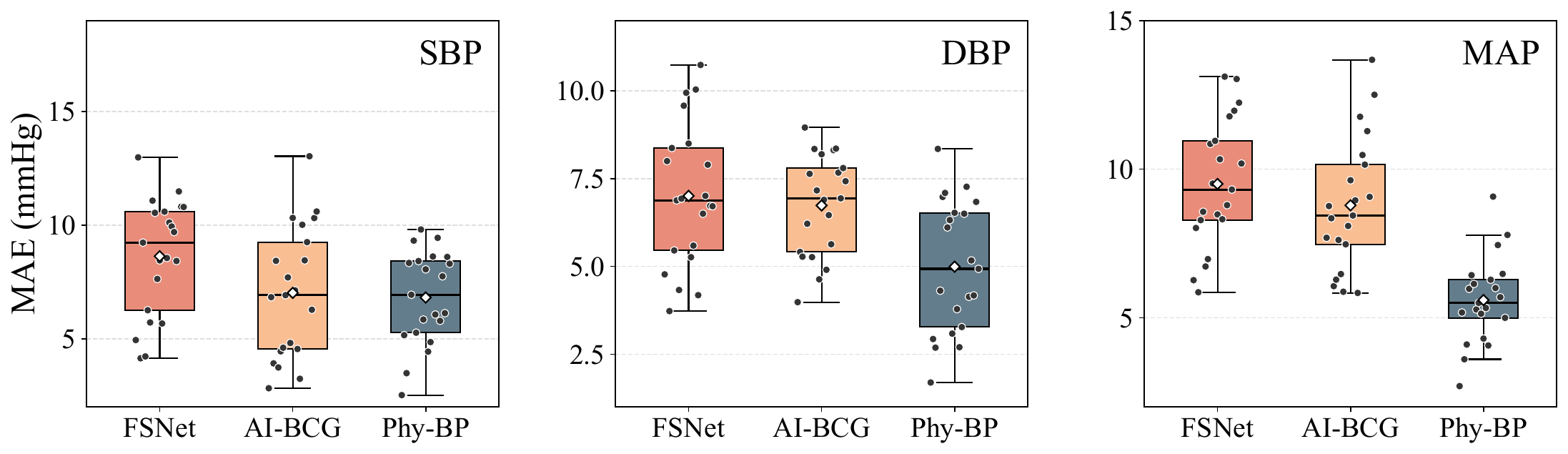}
            \put(0,29.1){\makebox(0,0)[tl]{{\footnotesize (a)}}}
        \end{overpic}}\\[3pt]
    \subfloat{\label{fig:subject_level_std}%
        \begin{overpic}[width=\columnwidth]{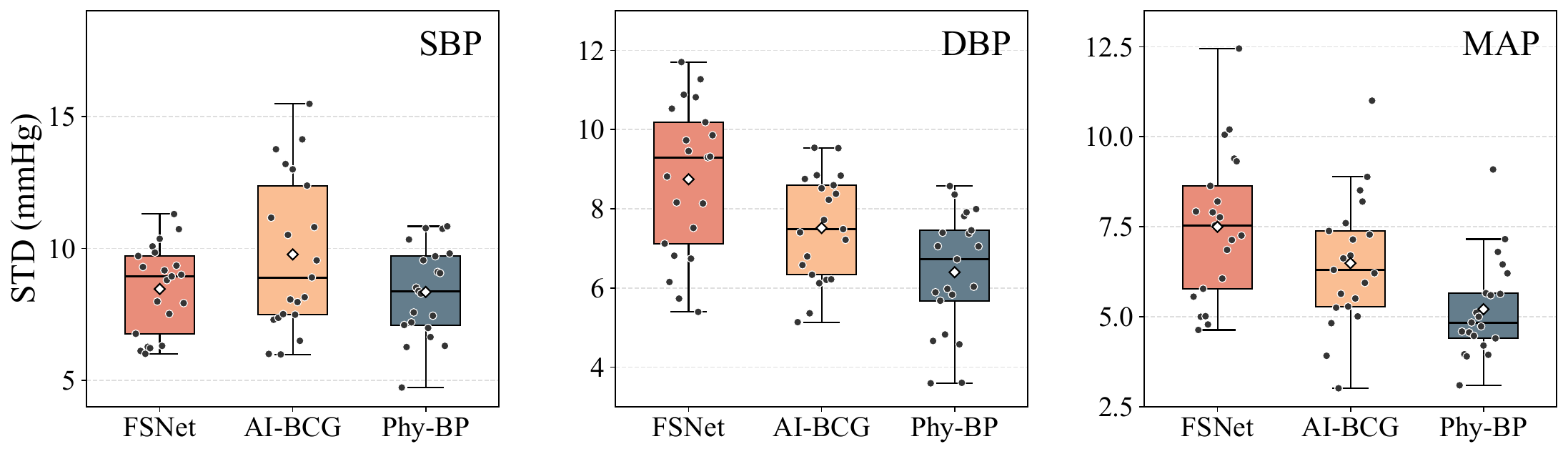}
            \put(0,29.1){\makebox(0,0)[tl]{{\footnotesize (b)}}}
        \end{overpic}}
    \caption{Subject-level BP estimation performance in terms of MAE and STD.}
    \label{fig:subject_level_metrics}
\end{figure}

\begin{table*}[!t]

\centering
\caption{Comparison results of different models.}
\label{tab:bp_results}
\setlength{\tabcolsep}{2pt}
\begin{adjustbox}{width=\textwidth}
\begin{tabular}{@{}l|cccc|cccc|cccc|r@{\,$\pm$\,}l|c@{}}
\toprule
\multirow{2}{*}{\textbf{Model}} 
& \multicolumn{4}{c|}{\textbf{SBP}} 
& \multicolumn{4}{c|}{\textbf{DBP}} 
& \multicolumn{4}{c|}{\textbf{MAP}} & \multicolumn{2}{c|}{\textbf{Overall}} & \multirow{2}{*}{\makecell[c]{\textbf{Subject-Level} \\ \textbf{Overall Gain}}} \\
& {MAE$\downarrow$} & {ME$\to$0} & {STD$\downarrow$} & {PCC$\uparrow$}
& {MAE$\downarrow$} & {ME$\to$0} & {STD$\downarrow$} & {PCC$\uparrow$}
& {MAE$\downarrow$} & {ME$\to$0} & {STD$\downarrow$} & {PCC$\uparrow$} & \multicolumn{2}{c|}{$\Delta m\%\uparrow$} &  \\
\midrule
\textbf{Hyde}~\cite{huang2026hyde} & $7.61$ & $1.23$ & $9.77$ & $0.73$ & $6.41$ & $1.05$ & $8.16$ & $0.65$ & $8.66$ & $1.81$ & $11.80$ & $0.67$ & \multicolumn{2}{c|}{baseline} & $35.85 \pm 6.24\%$ \\
\textbf{PITN}~\cite{wang2025pitn} & $12.82$ & $4.21$ & $15.59$ & $0.28$ & $7.44$ & $0.86$ & $8.91$ & $0.31$ & $9.10$ & $1.71$ & $11.20$ & $0.59$ & $-31.02$ & $2.12\%$ & $54.50 \pm 6.89\%$ \\
\textbf{MSTNN}~\cite{fan2023fewshotbp} & $12.60$ & $0.55$ & $15.58$ & $0.24$ & $6.98$ & $0.40$ & $9.36$ & $0.27$ & $9.56$ & $0.29$ & $12.29$ & $0.51$ & $-34.74$ & $3.36\%$ & $59.34 \pm 8.97\%$ \\
\textbf{FSNet}~\cite{zhang2026bcg} & $8.19$ & $0.79$ & $10.20$ & $0.75$ & $6.58$ & $-2.51$ & $7.85$ & $0.67$ & $8.25$ & $-2.91$ & $10.82$ & $0.66$ & $0.72$ & $1.77\%$ & $21.78 \pm 8.62\%$ \\
\textbf{AI-BCG}~\cite{huang2024ai} & $7.03$ & $0.56$ & $9.21$ & $0.77$ & $5.99$ & $0.34$ & $6.87$ & $0.65$ & $7.54$ & $0.04$ & $8.78$ & $0.79$ & $10.85$ & $2.92\%$ & $15.21 \pm 8.02\%$ \\
\midrule
\textbf{Phy-BP(3-axis)} & $6.03$ & $-0.05$ & $7.84$ & $0.89$ & $4.23$ & $-0.52$ & $5.90$ & $0.83$ & $5.07$ & $-0.33$ & $7.24$ & $0.86$ & $\mathbf{28.92}$ & $\mathbf{1.09\%}$ & -- \\
\bottomrule
\end{tabular}
\end{adjustbox}
\end{table*}

\subsubsection{Performance Overview}
\label{sec:benchmark_results}
Fig.~\ref{fig:result_phy} presents the Bland-Altman and correlation plots of the proposed Phy-BP for SBP, DBP and MAP estimation. As shown in Fig.~\ref{fig:result_phy}(a)--(c), the prediction errors are centered close to zero, with MEs of $-0.05$, $-0.52$ and $-0.33$~mmHg for SBP, DBP and MAP, respectively. On the current dataset, the corresponding STDs are $7.84$, $5.90$ and $7.24$~mmHg for SBP, DBP and MAP, respectively. The correlation plots in Fig.~\ref{fig:result_phy}(d)--(f) further show that the predictions follow the reference BP variations across subjects, with PCCs of $0.89$, $0.83$ and $0.86$ for SBP, DBP and MAP, respectively. These results suggest that Phy-BP does not merely predict a population-level average, but can track subject-dependent and time-varying BP changes in the hospital dataset.

The benchmark results in Table~\ref{tab:bp_results} show an overall improvement of $28.92\%$ for Phy-BP relative to the Hyde baseline. Phy-BP also achieves the lowest MAE and highest PCC across the compared models for all three BP targets. The Overall score is calculated from the reported sample-wise MAE, STD, and PCC values for SBP, DBP, and MAP relative to Hyde. The $\pm$ values in the Overall column denote the half-widths of $95\%$ confidence intervals across five complete LOSO evaluations on the same $21$ subjects. The Subject-Level Overall Gain column reports the average relative improvement of Phy-BP over each comparison model. Two-sided one-sample $t$-tests assess whether the mean improvement is zero, and all comparisons show significant improvements with $p < 0.05$. These results support the effectiveness of triaxial modeling and physics-constrained representation learning for BP estimation under the evaluated hospital conditions.

To illustrate subject-wise variability in BP estimation performance, Fig.~\ref{fig:subject_level_metrics} presents the MAE and STD distributions. These plots provide a complementary visual perspective on the distribution of performance across subjects. The spread of the distributions shows that some subjects remain challenging, with relatively high prediction errors. Nevertheless, Phy-BP generally achieves lower MAE and STD values across the three BP targets, indicating favorable overall performance despite this subject-wise variability.

It is worth noting that most existing BP estimation datasets are collected under controlled laboratory settings, where body posture, sensor attachment and signal morphology are relatively stable. In contrast, our dataset is naturally collected in a hospital scenario with routine patient movements, varying body-bed interactions and medication-induced hemodynamic changes. Under these more practical conditions, the compared models still show limited performance even after calibration, suggesting that the performance gap is not caused only by insufficient personalization but also by the mismatch between their signal assumptions and natural clinical deployment conditions.

\subsubsection{Comparisons to Models using Single-axis Inputs}
In Table~\ref{tab:bp_results}, Hyde serves as the baseline, while FSNet and AI-BCG achieve overall improvements of $0.72\%$ and $10.85\%$, respectively. These three single-axis BCG models outperform the PPG/ECG-based methods in this evaluation. However, single-axis sensing captures only a partial projection of the body-bed response induced by cardiogenic excitation. The performance gap relative to Phy-BP suggests that changes in network architecture alone may not fully compensate for this incomplete observation. Phy-BP combines triaxial BSG signals with physics-guided representation learning and achieves an overall improvement of $28.92\%$, supporting the value of incorporating the body-bed physical model and integrating complementary information across axes. Section~\ref{sec:triaxis_adv} further examines the contributions of different input axes through ablation studies.

\begin{figure}[t]
    \centering
    \captionsetup[subfloat]{farskip=0pt,nearskip=0pt}
    \subfloat{\label{fig:MAP_MAE}%
        \begin{overpic}[width=0.99\columnwidth]{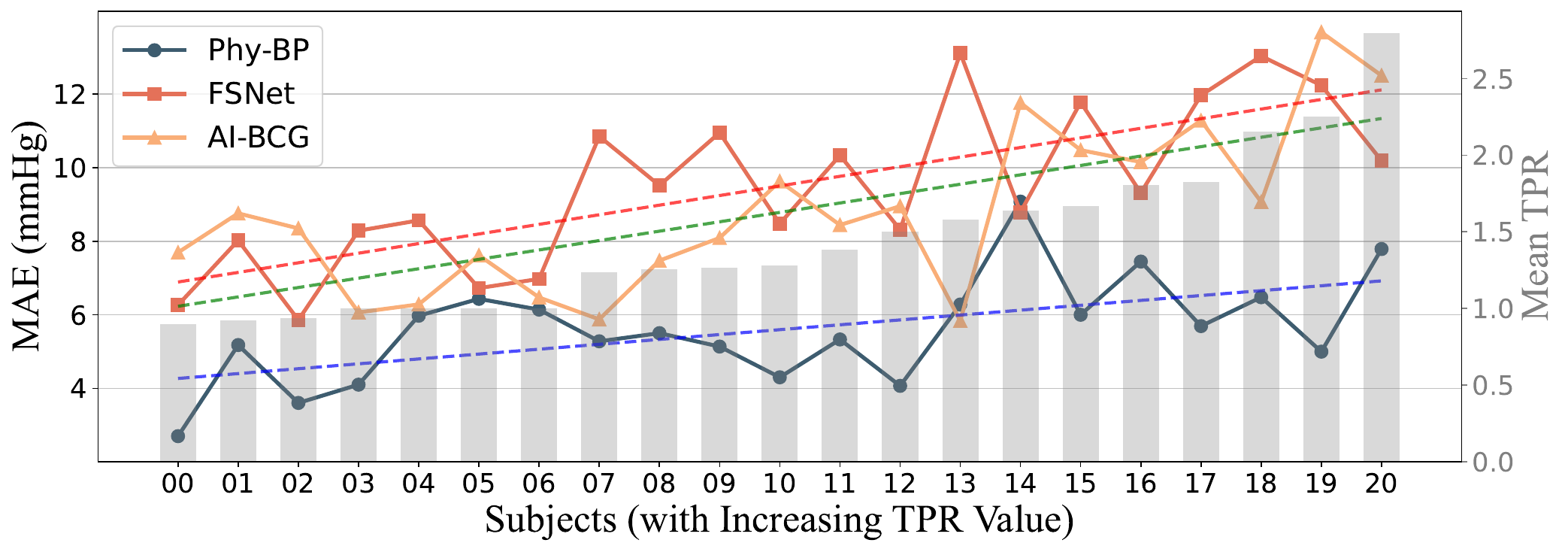}
            \put(0,35.2){\makebox(0,0)[tl]{{\footnotesize (a)}}}
        \end{overpic}}\\[3pt]
    \makebox[\columnwidth][l]{%
        \hspace*{0.006262\columnwidth}%
        \subfloat{\label{fig:tpr_unadjusted}%
            \begin{overpic}[width=0.45\columnwidth]{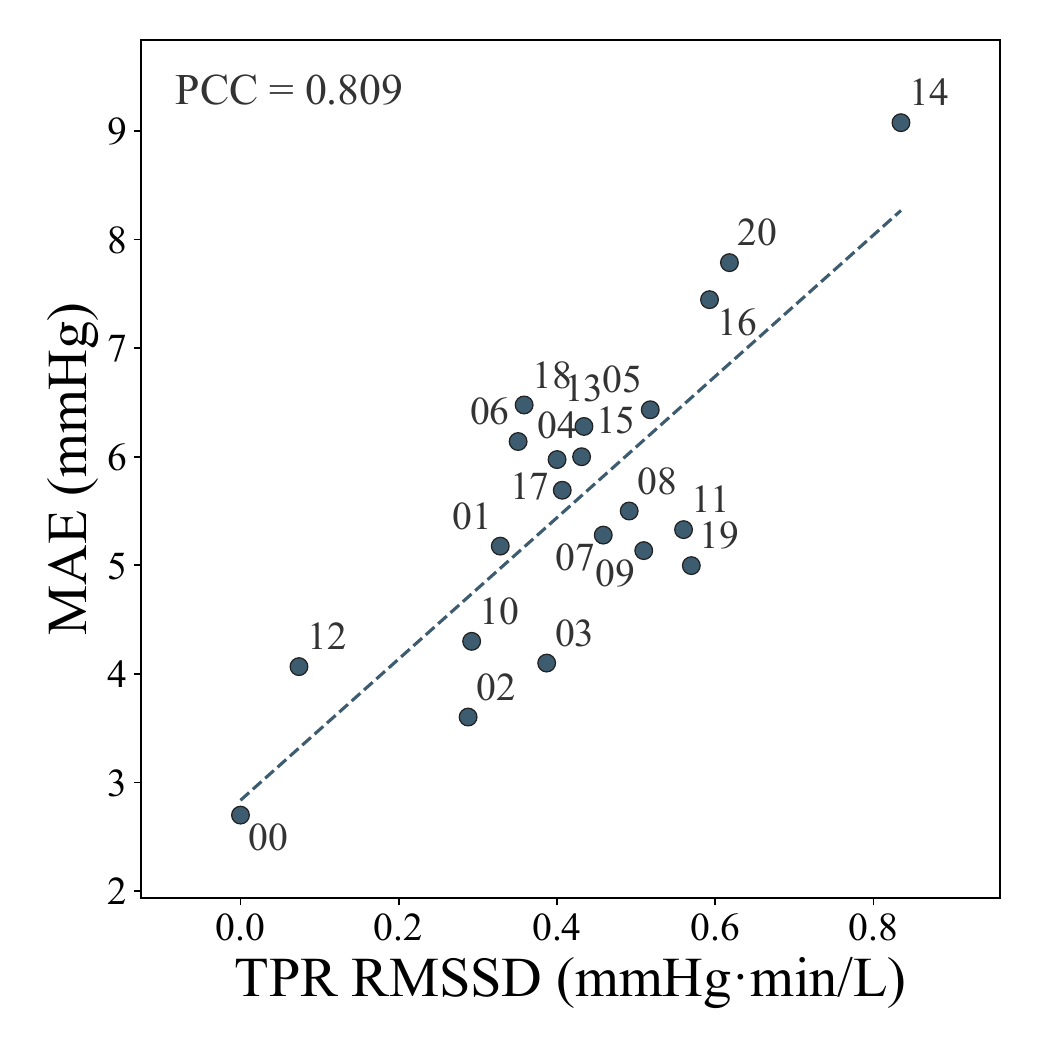}
                \put(-0.28052,100){\makebox(0,0)[tl]{{\footnotesize (b)}}}
            \end{overpic}}%
        \hspace*{0.037710\columnwidth}%
        \raisebox{-0.000472\columnwidth}{%
            \subfloat{\label{fig:tpr_adjusted}%
                \begin{overpic}[width=0.450853\columnwidth]{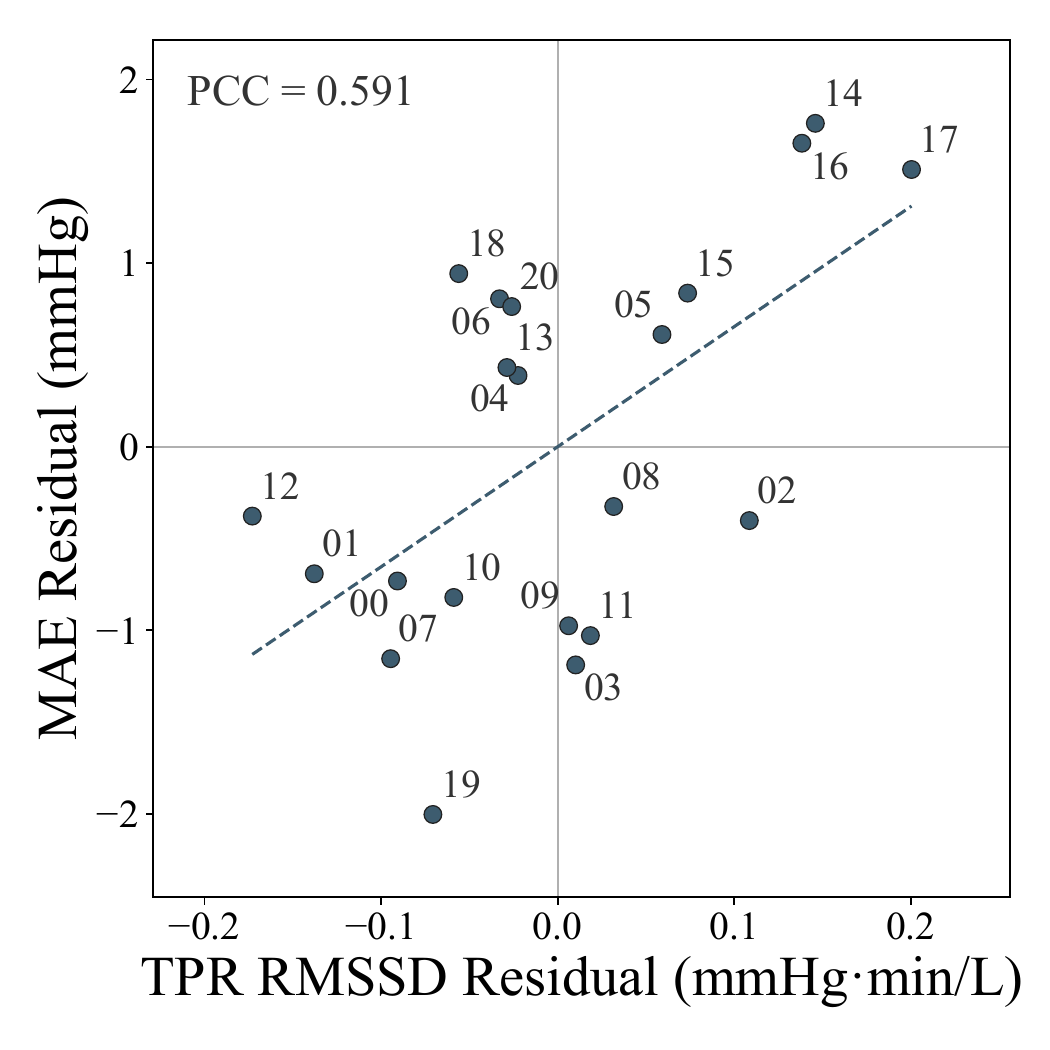}
                    \put(0,99.91548){\makebox(0,0)[tl]{{\footnotesize (c)}}}
                \end{overpic}}}}
    \caption{MAP prediction error across subjects: (a) MAE of MAP prediction; (b) and (c) Association between TPR variability and MAP prediction error before and after adjustment for MAP RMSSD, respectively.}
    \label{fig:tpr}
\end{figure}
\subsubsection{Analysis for Different TPR}
To further examine the impact of subject-specific hemodynamic variation, we analyze MAP prediction performance with respect to TPR variability. Most previous cuffless BP studies lack CO measurements to estimate TPR, leaving it unclear whether their models learn transferable cardiovascular representations or rely on stable subject-specific hemodynamics. In our dataset, TPR is calculated from reference MAP and PiCCO-derived CO. Fig.~\ref{fig:tpr}(a) compares Phy-BP, FSNet, and AI-BCG across subjects ordered by increasing mean TPR. Phy-BP generally achieves lower MAP MAE than the other two models across subjects.

Across the $21$ subjects, TPR variability, measured by the root mean square of successive differences (RMSSD), has an unadjusted PCC of $0.809$ with the MAP MAE of Phy-BP (Fig.~\ref{fig:tpr}(b)). To assess the association after accounting for reference-MAP variability, we regress both metrics separately on reference-MAP RMSSD using ordinary least squares with an intercept. The Pearson correlation between the residuals (observed minus fitted values) yields an adjusted partial correlation of $0.591$ (Fig.~\ref{fig:tpr}(c)).

The correlation decreases by $0.218$ after adjustment, showing that accounting for reference-MAP variability weakens the observed association. The remaining positive association suggests that greater TPR variability is associated with larger MAP prediction error. Several subjects deviate from the adjusted trend, potentially because MAP fluctuations contribute differently to derived TPR variability and prediction error across subjects. However, the adjustment does not quantify the exact contribution of mathematical coupling or eliminate all dependence on the shared MAP term. Explicit modeling of the relationship between TPR and BSG signals therefore remains an important direction for future work.

\subsubsection{Computational Complexity}\label{sec:complexity}
\begin{figure}[t]
    \centering
    \includegraphics[width=0.90\columnwidth]{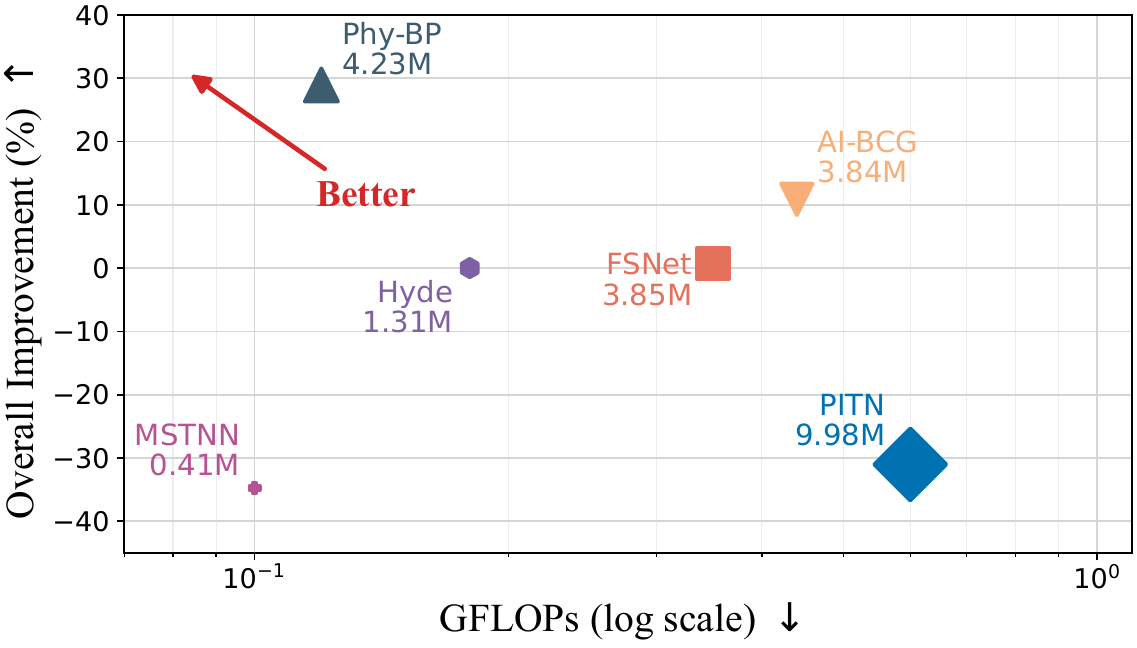}
    \caption{Overall performance versus computational complexity, with marker sizes representing parameter counts.}
    \label{fig:overall_complexity}
\end{figure}

Fig.~\ref{fig:overall_complexity} relates the overall improvement in Table~\ref{tab:bp_results} to computational cost and model size. The horizontal axis shows computational cost in billions of floating-point operations (GFLOPs), while marker sizes and annotations indicate parameter counts in millions. Compared with the other models, Phy-BP achieves the lowest MAE and highest PCC for all three BP targets with a moderate parameter count. This design reflects our rationale that effective feature extraction does not necessarily require larger encoders, and conventional ResNet encoders can instead learn useful representations when feature evolution is guided by the physical constraint derived from the body-bed system.

\subsubsection{Performance under Limited Training Dataset Scale}\label{sec:training_scale}
\begin{figure}[t]
    \centering
    \captionsetup[subfloat]{farskip=0pt,nearskip=0pt}
    \subfloat{\label{fig:mae_change}%
        \raisebox{0pt}[\dimexpr\height+3pt\relax][\depth]{%
        \begin{overpic}[width=0.5\columnwidth]{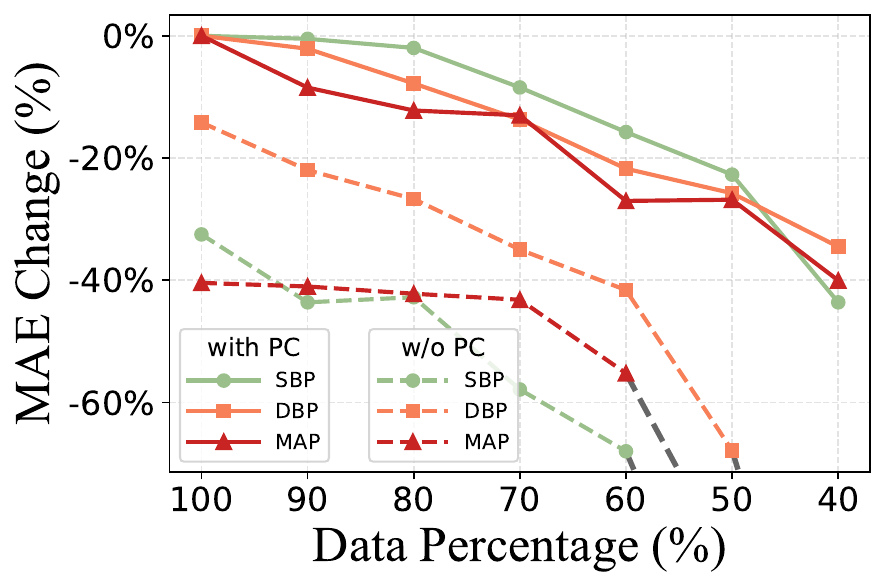}
            \put(0,68.78){\makebox(0,0)[tl]{{\footnotesize (a)}}}
        \end{overpic}}}%
    \subfloat{\label{fig:std_change}%
        \raisebox{0pt}[\dimexpr\height+3pt\relax][\depth]{%
        \begin{overpic}[width=0.5\columnwidth]{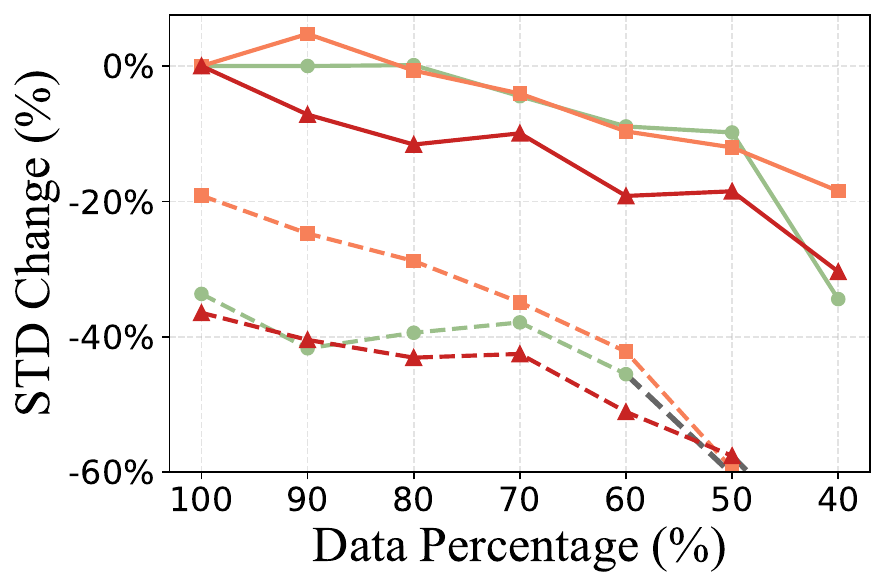}
            \put(0,68.78){\makebox(0,0)[tl]{{\footnotesize (b)}}}
        \end{overpic}}}\\[3pt]
    \subfloat{\label{fig:pcc_change}%
        \raisebox{0pt}[\dimexpr\height+3pt\relax][\depth]{%
        \begin{overpic}[width=0.5\columnwidth]{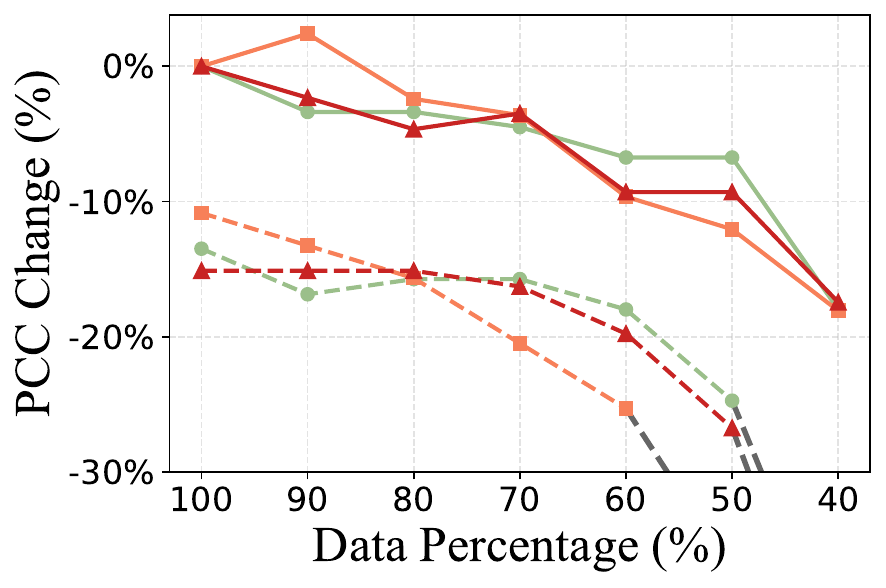}
            \put(0,68.78){\makebox(0,0)[tl]{{\footnotesize (c)}}}
        \end{overpic}}}%
    \subfloat{\label{fig:overall_change}%
        \raisebox{0pt}[\dimexpr\height+3pt\relax][\depth]{%
        \begin{overpic}[width=0.5\columnwidth]{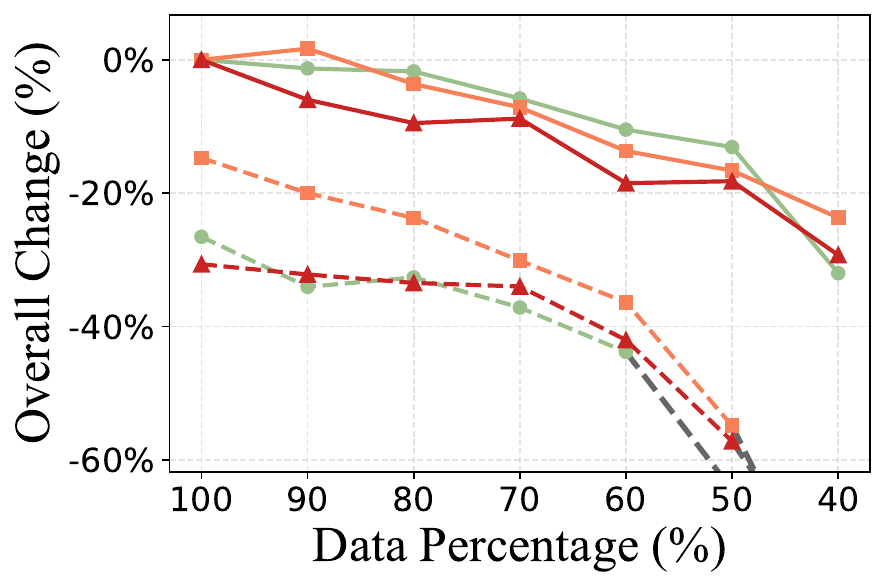}
            \put(0,68.78){\makebox(0,0)[tl]{{\footnotesize (d)}}}
        \end{overpic}}}
    \caption{Training Phy-BP using different scales of data: (a) - (c) MAE, STD and PCC changes for the model with or without the physics constraint (PC); (d) Overall degradation in terms of SBP, DBP and MAP.}
    \label{fig:data_scale}
\end{figure}

Fig.~\ref{fig:data_scale} compares performance at reduced training scales relative to Phy-BP trained on $100\%$ of the data with the physics constraint ($0\%$ change). MAE, STD, and PCC deteriorate modestly with moderate data reductions but more markedly at smaller scales. Without the physics constraint, prediction performance degrades substantially at smaller training scales. The low-performance results are shown as dashed gray lines in Fig.~\ref{fig:data_scale}, while some severely degraded results with little practical meaning are omitted from the displayed ranges for clarity. With the physics constraint, Phy-BP maintains reasonable performance using only $40\%$ of the training data. The shared dynamical structure among triaxial BSG representations helps the network learn consistent cardiogenic features from limited samples, supporting the data efficiency and robustness of physics-guided regularization.

\subsection{Ablation Studies }~\label{sec:ablation_studies}
\begin{figure}[t]
    \centering
    \subfloat[Cycles$>4$]{\label{fig:quality_0}\includegraphics[width=0.5\columnwidth]{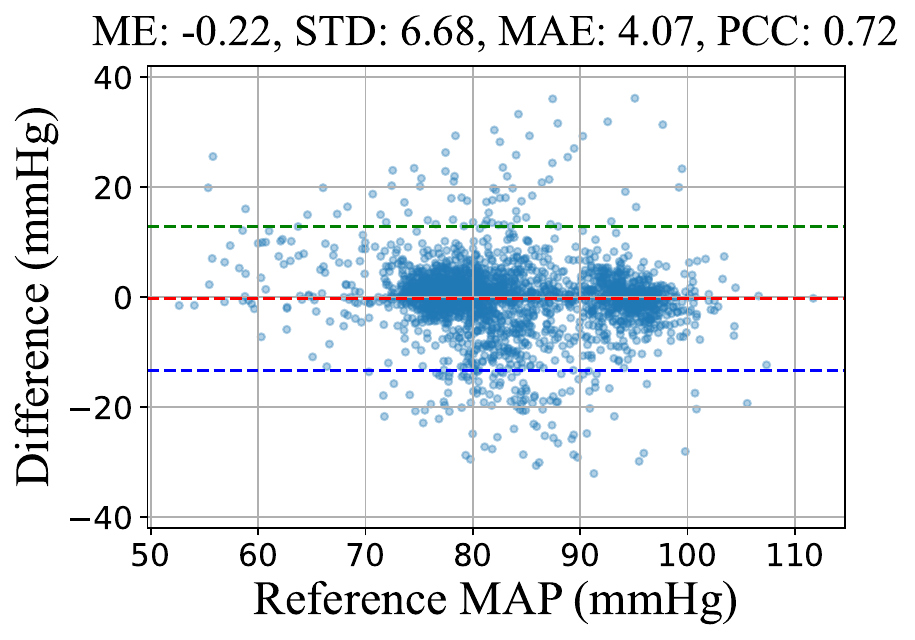}}
    \subfloat[Cycles$>5$]{\label{fig:quality_1}\includegraphics[width=0.5\columnwidth]{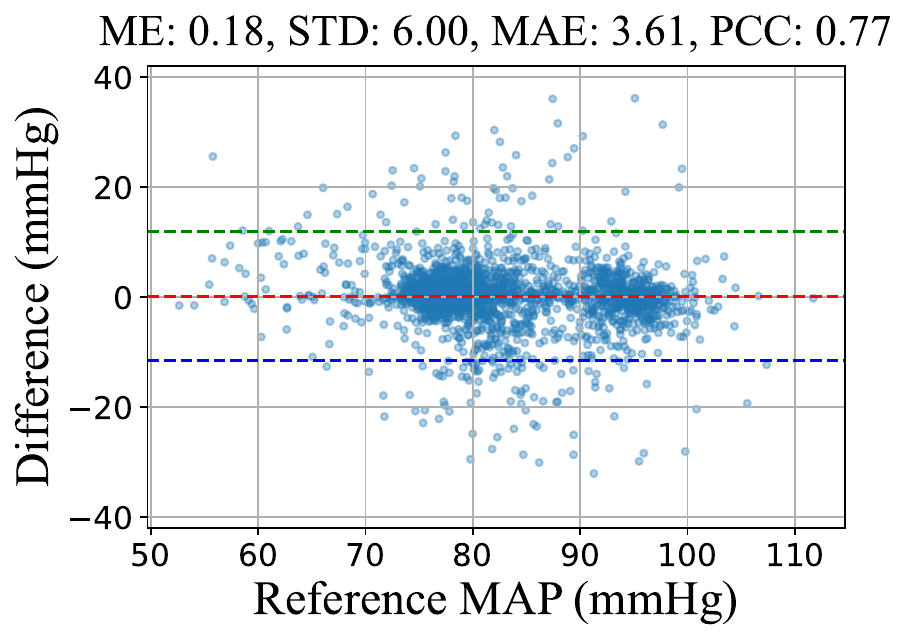}} \\
    \subfloat[Cycles$>6$]{\label{fig:quality_2}\includegraphics[width=0.5\columnwidth]{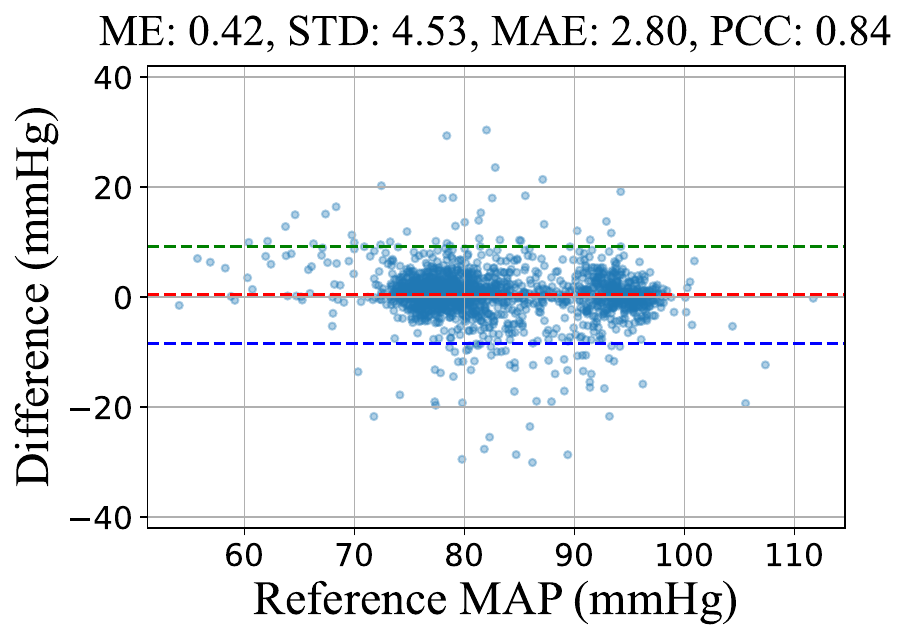}} 
    \subfloat[Cycles$>7$]{\label{fig:quality_3}\includegraphics[width=0.5\columnwidth]{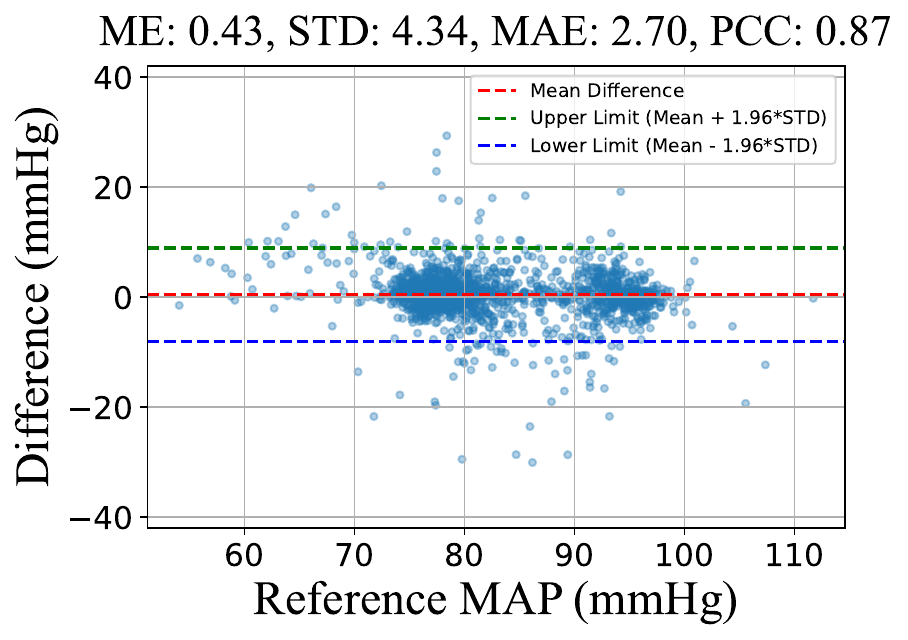}} 
    \caption{Effectiveness of the quality control with different thresholds for required matched cardiac cycles.}
    \label{fig:qctrl}
\end{figure}
\subsubsection{Effectiveness of Quality Control}
\begin{figure*}[t]
    \centering
    \captionsetup[subfloat]{farskip=0pt,nearskip=0pt}
    \subfloat{\label{fig:quality_MAE}%
                \begin{overpic}[width=0.65\columnwidth]{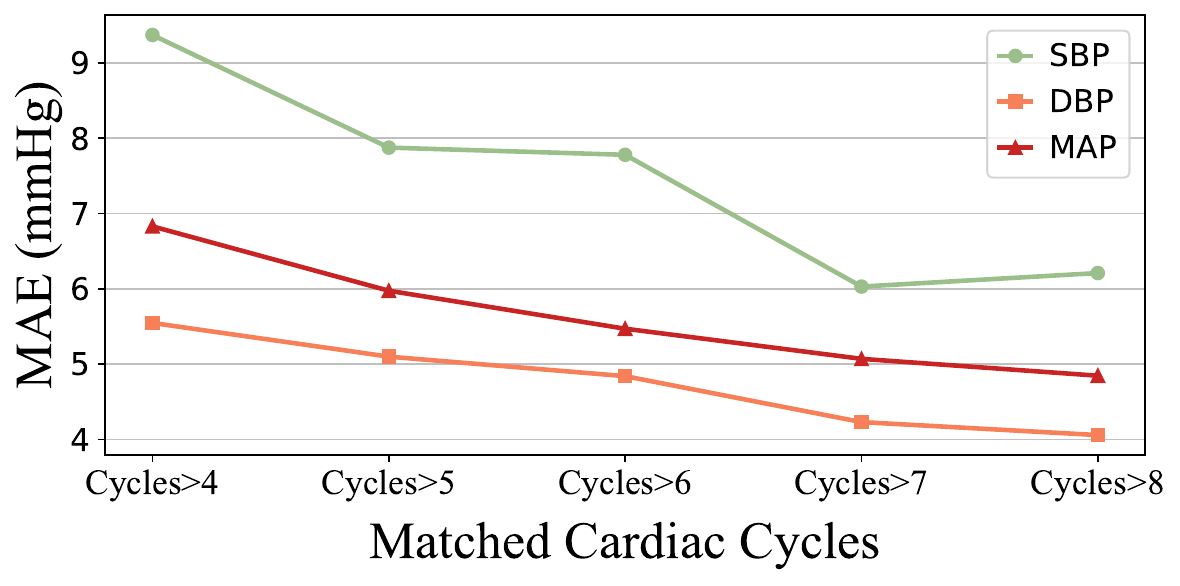}
            \put(0,49.59){\makebox(0,0)[tl]{{\footnotesize (a)}}}
        \end{overpic}}\hfill
    \subfloat{\label{fig:quality_STD}%
                \begin{overpic}[width=0.65\columnwidth]{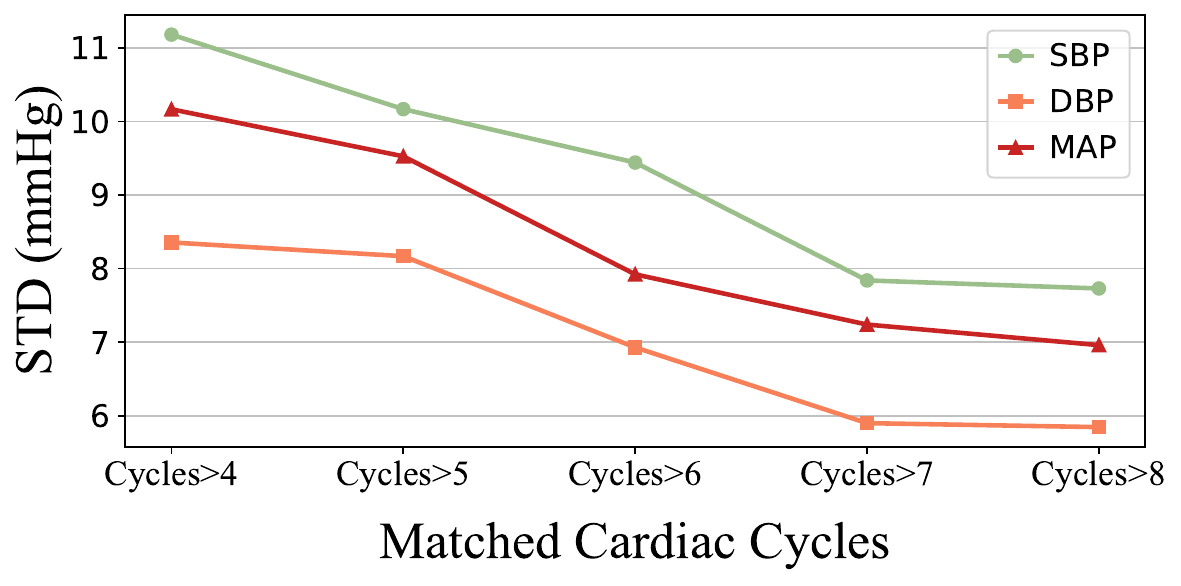}
            \put(0,49.55){\makebox(0,0)[tl]{{\footnotesize (b)}}}
        \end{overpic}}\hfill
    \subfloat{\label{fig:quality_PCC}%
                \begin{overpic}[width=0.65\columnwidth]{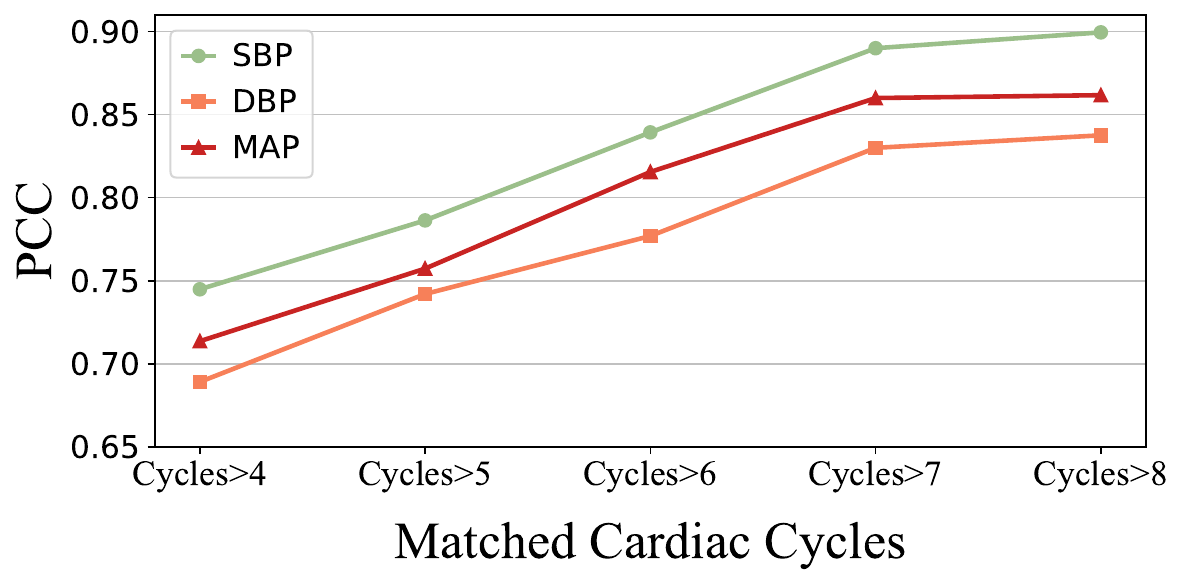}
            \put(0,49.45){\makebox(0,0)[tl]{{\footnotesize (c)}}}
        \end{overpic}}\\[3pt]
    \subfloat{\label{fig:qc_retention}%
        \begin{overpic}[width=0.98\columnwidth]{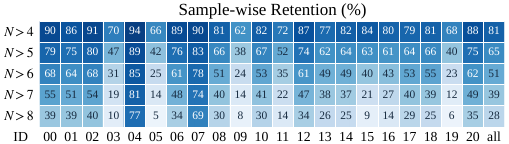}
            \put(0,29){\makebox(0,0)[tl]{{\footnotesize (d)}}}
        \end{overpic}}\hfill
    \subfloat{\label{fig:qc_hourly_coverage}%
        \begin{overpic}[width=0.98\columnwidth]{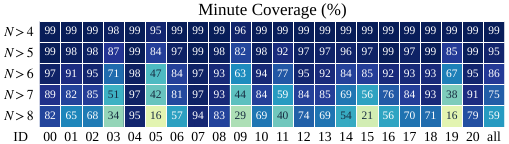}
            \put(0,29){\makebox(0,0)[tl]{{\footnotesize (e)}}}
        \end{overpic}}
    
    \caption{Effectiveness and coverage after quality control: (a) - (c) The performance of BP prediction in terms of MAE, STD and PCC, respectively; (d) - (e) Sample-wise retention and minute coverage with different quality-control thresholds.}
    \label{fig:qctrl_perf}
\end{figure*}

Fig.~\ref{fig:qctrl} illustrates the effect of quality control for one subject. As the matched-cycle threshold increases from Cycles$>4$ to Cycles$>7$, large-error samples are progressively removed, while the core cluster of accurate predictions remains near the zero-error line. Across the dataset, applying stricter thresholds during inference with the same trained model generally reduces MAE and STD and improves PCC on retained segments, as shown in Fig.~\ref{fig:qctrl_perf}(a)--(c). These results support the use of matched-cycle counts to identify BSG segments containing reliable cardiogenic information for BP estimation.

Stricter filtering reduces data availability, so the threshold must balance signal reliability and monitoring coverage. We use Cycles $>7$, requiring at least eight matched cardiac cycles per $10$-s segment. Fig.~\ref{fig:qctrl_perf}(d) and Fig.~\ref{fig:qctrl_perf}(e) show sample-wise retention and minute coverage, respectively. Minute coverage is the proportion of recording minutes with at least one retained segment among minutes containing data before filtering. At this threshold, mean subject-level retention is $39.24\%$, pooled retention is $37.61\%$, and minute coverage is approximately $75\%$. Retained segments therefore cover most recording minutes, although many segments are excluded. For monitoring tasks that do not require frequent BP estimation, reduced sample-wise retention can coexist with relatively high temporal monitoring coverage.

\subsubsection{Necessity of using Triaxial BSG for BP Monitoring}\label{sec:triaxis_adv}
\begin{table*}[!t]
\centering

\caption{Ablation studies under different settings}
\label{tab:bp_tri}
\resizebox{\textwidth}{!}{%
\begin{tabular}{c|cccc|cccc|cccc|c}
\toprule 
{\textbf{Settings for}} 
& \multicolumn{4}{c|}{\textbf{SBP}} 
& \multicolumn{4}{c|}{\textbf{DBP}} 
& \multicolumn{4}{c|}{\textbf{MAP}} &{\textbf{Overall}} \\
\textbf{Phy-BP}& {MAE$\downarrow$} & {ME$\to$0} & {STD$\downarrow$} & {PCC$\uparrow$}
& {MAE$\downarrow$} & {ME$\to$0} & {STD$\downarrow$} & {PCC$\uparrow$}
& {MAE$\downarrow$} & {ME$\to$0} & {STD$\downarrow$} & {PCC$\uparrow$} & $\Delta m\%\uparrow$\\
\midrule
  & \multicolumn{12}{c}{\textbf{Using Different Input Axes}} \\ 
\midrule
\textbf{Y only} & $10.64$ & $5.19$ & $15.63$ & $0.41$ & $6.65$ & $0.99$ & $7.71$ & $0.44$ & $6.28$ & $1.86$ & $8.79$ & $0.65$ & $-13.80\%$ \\
\textbf{Z only} & $10.28$ & $-1.20$ & $11.96$ & $0.70$ & $5.58$ & $-0.27$ & $8.62$ & $0.72$ & $7.11$ & $-1.59$ & $10.41$ & $0.73$ & $-0.54\%$ \\
\textbf{X+Y} & $6.32$ & $-1.14$ & $7.23$ & $0.78$ & $5.12$ & $0.22$ & $7.16$ & $0.72$ & $6.79$ & $0.13$ & $9.78$ & $0.76$ & $16.12\%$ \\
\textbf{X+Z} & $8.37$ & $-1.66$ & $10.03$ & $0.68$ & $6.32$ & $0.85$ & $8.39$ & $0.61$ & $6.40$ & $0.59$ & $11.63$ & $0.62$ & $-0.78\%$ \\
\textbf{Y+Z} & $9.29$ & $0.19$ & $10.24$ & $0.83$ & $4.94$ & $0.20$ & $7.20$ & $0.74$ & $5.97$ & $0.19$ & $8.57$ & $0.82$ & $12.91\%$ \\
\textbf{X+Y+Z} & $6.03$ & $-0.05$ & $7.84$ & $0.89$ & $4.23$ & $-0.52$ & $5.90$ & $0.83$ & $5.07$ & $-0.33$ & $7.24$ & $0.86$ & $\mathbf{28.92\%}$ \\
\midrule
  & \multicolumn{12}{c}{\textbf{Using Different Fusion Configurations}} \\
\midrule
\textbf{Mean fusion} & $7.83$ & $1.03$ & $9.87$ & $0.72$ & $5.70$ & $-1.53$ & $7.65$ & $0.71$ & $7.67$ & $-0.86$ & $10.28$ & $0.70$ & $5.56\%$ \\
\textbf{Gate only} & $7.99$ & $-1.32$ & $10.48$ & $0.77$ & $4.83$ & $-0.12$ & $7.03$ & $0.74$ & $7.12$ & $-0.54$ & $9.88$ & $0.73$ & $9.84\%$ \\
\textbf{Physics only} & $6.57$ & $-0.23$ & $8.51$ & $0.83$ & $5.05$ & $-2.29$ & $5.08$ & $0.81$ & $5.58$ & $-1.77$ & $7.93$ & $0.84$ & $24.17\%$ \\
\textbf{Channel Atten.}$^{*}$ & $7.49$ & $0.35$ & $8.27$ & $0.70$ & $5.75$ & $-2.37$ & $7.77$ & $0.75$ & $6.60$ & $-1.72$ & $10.31$ & $0.73$ & $9.85\%$ \\
\textbf{Cross-axis Atten.}$^{*}$ & $6.83$ & $-0.47$ & $8.05$ & $0.84$ & $5.03$ & $-1.23$ & $6.54$ & $0.80$ & $5.00$ & $-1.69$ & $7.02$ & $0.86$ & $24.28\%$ \\
\textbf{Full Model} & $6.03$ & $-0.05$ & $7.84$ & $0.89$ & $4.23$ & $-0.52$ & $5.90$ & $0.83$ & $5.07$ & $-0.33$ & $7.24$ & $0.86$ & $\mathbf{28.92\%}$ \\
\midrule
  & \multicolumn{12}{c}{\textbf{Using Different Weighting Coefficient $\lambda$}} \\ 
\midrule
\textbf{$\mathbf{\lambda=0.0}$} & $7.99$ & $-1.32$ & $10.48$ & $0.77$ & $4.83$ & $-0.12$ & $7.03$ & $0.74$ & $7.12$ & $-0.54$ & $9.88$ & $0.73$ & $9.84\%$ \\
\textbf{$\mathbf{\lambda=0.5}$} & $9.40$ & $-1.54$ & $11.02$ & $0.85$ & $4.97$ & $-0.15$ & $7.01$ & $0.79$ & $6.84$ & $0.47$ & $10.42$ & $0.82$ & $10.37\%$ \\
\textbf{$\mathbf{\lambda=1.0}$} & $8.71$ & $-1.45$ & $9.49$ & $0.85$ & $5.18$ & $0.64$ & $6.69$ & $0.78$ & $6.24$ & $0.59$ & $9.38$ & $0.85$ & $15.26\%$ \\
\textbf{$\mathbf{\lambda=1.5}$} & $6.03$ & $-0.05$ & $7.84$ & $0.89$ & $4.23$ & $-0.52$ & $5.90$ & $0.83$ & $5.07$ & $-0.33$ & $7.24$ & $0.86$ & $\mathbf{28.92\%}$ \\
\textbf{$\mathbf{\lambda=2.0}$} & $7.01$ & $-0.33$ & $9.30$ & $0.83$ & $4.42$ & $-1.15$ & $6.48$ & $0.80$ & $5.83$ & $-1.26$ & $8.31$ & $0.81$ & $20.47\%$ \\
\textbf{$\mathbf{\lambda=2.5}$} & $7.48$ & $0.15$ & $8.51$ & $0.83$ & $4.82$ & $-0.59$ & $7.09$ & $0.76$ & $7.03$ & $0.08$ & $9.73$ & $0.85$ & $16.26\%$ \\
\midrule
  & \multicolumn{12}{c}{\textbf{Using Different Calibration Intervals}} \\ 
\midrule
\textbf{No Calibration} & $15.54$ & $-0.19$ & $20.05$ & $0.56$ & $7.56$ & $-1.29$ & $9.91$ & $0.41$ & $9.77$ & $-0.84$ & $12.39$ & $0.55$ & $-38.31\%$\\
\textbf{1st Sample Only} & $9.29$ & $0.22$ & $10.07$ & $0.79$ & $4.94$ & $0.70$ & $6.98$ & $0.80$ & $6.81$ & $0.23$ & $9.89$ & $0.74$ & $10.17\%$ \\
\textbf{Every 40 h} & $9.31$ & $-0.95$ & $10.48$ & $0.82$ & $4.38$ & $-0.29$ & $6.84$ & $0.79$ & $6.18$ & $-0.39$ & $9.19$ & $0.78$ & $13.25\%$ \\
\textbf{Every 32 h} & $8.76$ & $0.76$ & $9.84$ & $0.80$ & $4.11$ & $-1.69$ & $6.19$ & $0.80$ & $5.86$ & $-0.14$ & $8.57$ & $0.82$ & $17.66\%$ \\
\textbf{Every 16 h} & $8.86$ & $0.13$ & $9.86$ & $0.85$ & $4.04$ & $-0.67$ & $5.97$ & $0.85$ & $5.12$ & $-0.33$ & $8.17$ & $0.86$ & $21.52\%$\\
\textbf{Every \phantom{0}8 h} & $6.03$ & $-0.05$ & $7.84$ & $0.89$ & $4.23$ & $-0.52$ & $5.90$ & $0.83$ & $5.07$ & $-0.33$ & $7.24$ & $0.86$ & $\mathbf{28.92\%}$ \\
\bottomrule
\end{tabular}%
}
\par\vspace{2pt}{\footnotesize\raggedright\noindent $^{*}$ In these variants, channel attention or cross-axis attention replaces the gate-based fusion in the Full Model.\par}
\end{table*}

Table~\ref{tab:bp_tri} evaluates different combinations of BSG axes and shows that vibration responses along different directions provide complementary information for BP estimation. Adding the X-axis to the conventional Y-axis input increases the Overall score from $-13.80\%$ to $16.12\%$, consistent with the energy leakage illustrated in Fig.~\ref{fig:bsg_quality}(b), where cardiogenic vibration appears in directions not fully captured by Y-axis sensing. The Z-axis alone also substantially outperforms the Y-axis and achieves performance close to the BCG-based methods in Table~\ref{tab:bp_results}, supporting the use of Z-axis sensing in existing BCG studies. Combining Y and Z achieves an overall improvement of $12.91\%$, below the $16.12\%$ achieved by X+Y, while using all three axes achieves the best result of $28.92\%$. These results support triaxial monitoring for capturing a more complete body-bed response to cardiogenic excitation and improving BP estimation.

\subsubsection{Contributions of Triaxial Fusion Components}\label{sec:fusion_ablation}
Table~\ref{tab:bp_tri} separates the contributions of gate-based alignment and physics-guided regularization using four configurations: Mean fusion, Gate only, Physics only, and the Full Model. Mean fusion directly averages the triaxial features, the two single-component variants enable gate-based alignment or the physical constraint separately, and the Full Model combines both. All configurations use the same triaxial inputs and encoder and decoder architectures. Physics only achieves an Overall score of 24.17\%, close to 28.92\% for the Full Model and higher than either Mean fusion or Gate only. These results suggest that constraining latent feature evolution is the main contributor in these ablations, while gate-based alignment provides a further improvement when combined with physics-guided regularization.

Channel attention and cross-axis attention replace gate-based fusion while retaining the physical constraint, evaluating adaptive feature-channel weighting and inter-axis interactions, respectively. Channel attention reduces overall performance, possibly because channel reweighting suppresses weaker but complementary cardiac features. Cross-axis attention achieves similar MAP accuracy but provides no consistent advantage over the Full Model. In the gate-based design, the arithmetic-mean prototype serves as a shared reference, while time-varying gates adaptively balance each axis feature against this reference. These comparisons support retaining simple gate-based fusion to complement physics-guided representation learning, as the additional attention computations provide limited benefit and are less aligned with our future aim of real-time prediction.

\subsubsection{Effects of Physics Constraint}
Table~\ref{tab:bp_tri} evaluates the effect of the physical constraint by varying the weighting coefficient $\lambda$. The model already benefits from complementary triaxial BSG information without physics-guided regularization. This agrees with the input-axis ablation, as observing multiple vibration directions provides a more complete description of the body-bed response to cardiogenic excitation. Introducing the physical constraint further improves performance, achieving the best overall improvement of $28.92\%$. The gains are mainly reflected in lower STD and higher PCC for BP. These results suggest that physics-guided regularization promotes more consistent multi-axis representations, reducing prediction error dispersion and better preserving BP variation. The physical constraint therefore contributes to the reliability and consistency of BP prediction.

\subsubsection{Effects of Calibration}
Calibration is an indispensable stage for clinical BP monitoring because cuffless and contactless models cannot directly observe the subject-specific hemodynamic condition, especially the individual TPR. The need for subsequent reference updates depends on the hemodynamic stability of the deployment scenario and the required accuracy. The last part of Table~\ref{tab:bp_tri} presents results without calibration, with only the first calibration sample, and with periodic calibration every $40$, $32$, $16$, or $8$ hours. Without calibration, the model suffers from large errors and even shows a negative overall improvement of $-38.31\%$, indicating that a population-level model alone is insufficient for patients with unstable hemodynamic states. Calibration every $40$ hours yields an overall score of $13.25\%$ and lower MAE/STD for all BP targets than the uncalibrated setting. This result suggests that infrequent reference measurements can effectively correct the model response in hospital monitoring, where patients often experience dynamic hemodynamic changes.

Table~\ref{tab:bp_tri} also evaluates sensitivity to the calibration interval. With only the first subject-specific reference sample and no later updates, the Overall score is $10.17\%$, and the pooled PCCs across evaluated segments are $0.79$, $0.80$, and $0.74$ for SBP, DBP, and MAP, respectively. Extending the interval generally reduces overall prediction accuracy, and the feasibility of a longer interval depends on the precision and temporal requirements of the monitoring task. Initial subject-specific calibration remains necessary for reliable absolute BP estimation in the present framework, but this dependence does not imply that frequent recalibration is essential for every task. Calibration every $16$ hours still maintains relatively low bias and error dispersion, with MAP ME and STD of $-0.33$ and $8.17$~mmHg, respectively. More frequent updates may therefore be unnecessary for tasks that can tolerate these error levels. The calibration schedule should balance the required accuracy against the measurement burden and should be adapted to individual hemodynamic stability.

\subsection{Limitations and Future Work}\label{sec:limitations}
\subsubsection{Physiological Modeling and Calibration}\label{sec:limitation_physiology}
The current physical constraint models mechanical propagation through the body-bed system, while the mapping from cardiogenic motion to BP also depends on TPR and other physiological states. These factors are not explicitly represented by the reduced model, and subject-specific reference calibration remains necessary in the present framework. Before the first subject-specific calibration, the model produces estimates based on the relationships learned from the training population. These outputs may reflect relative BP changes, but their absolute levels have not been adjusted using an individual reference measurement.

Existing studies have shown correlations between features of human cardiogenic vibration signals and aortic pulse pressure~\cite{kim2016mechanism}, cardiac output~\cite{zhan2024cardiac}, and diastolic function~\cite{agam2022diastolic}. However, these correlations alone do not establish a unique quantitative mapping from vibration signals to absolute BP or TPR. The underlying mechanisms and their dependence on individual physiological conditions remain incompletely understood. Future work will develop theoretical models that couple mechanical dynamics with hemodynamic regulation to clarify these relationships, examine the roles of TPR and other physiological states, and identify the information needed to improve absolute BP estimation and reduce calibration requirements. Building on these models, we will investigate incorporating cardiovascular fluid dynamics priors into the loss function or decoding network to strengthen the physiological basis of BP estimation.

\subsubsection{Validation Across Beds and Mattresses}\label{sec:limitation_beds}
The present evaluation uses the same type of hospital bed, and the applicability of the reduced model across different beds and mattresses remains to be established. The second-order approximation provides a compact structure for latent dynamics, but its adequacy under substantially different mechanical conditions or strongly nonlinear and time-varying responses requires further investigation. Future work will expand data collection across different bed structures and mattress materials and examine the effects of body posture, body weight, and sensor placement. Comparing waveform-fitting quality, modal characteristics, and BP estimation performance across these conditions will help determine when the current reduced structure remains suitable and when its parameters or model form require adaptation.

\subsubsection{Abnormal Cardiac Rhythms}\label{sec:limitation_quality}
At the default quality-control threshold, ECG-flagged segments accounted for $2.55\%$ of all segments before filtering and approximately $2.37\%$ of retained segments afterward. With a pooled segment retention rate of $37.61\%$, these proportions indicate an estimated retention rate of approximately $34.96\%$ for ECG-flagged segments. The template-based quality-control procedure may exclude physiologically meaningful BSG cycles whose morphology differs from the reference templates, including cycles recorded during arrhythmias. BP estimation during arrhythmias requires further validation. Future work will collect synchronized BSG, ECG, and reference BP recordings during arrhythmias, with independent annotations of rhythm and signal quality to distinguish physiological waveform changes from recording artifacts and evaluate BP estimation under these conditions.

\subsubsection{Balance Between Monitoring Coverage and Accuracy}\label{sec:limitation_coverage}
The matched-cycle threshold aims to retain sufficient repeated cardiogenic information for BP estimation, but stricter requirements can reduce segment retention and monitoring coverage. The empirical balance in the present dataset may not generalize to other populations or recording conditions. In particular, a fixed count within a $10$-sec segment may limit retention at lower heart rates, even when individual cardiac cycles remain recognizable. Consequently, prediction errors on retained segments alone do not fully characterize monitoring performance. Future work will investigate single-cycle morphological confidence scores and the joint analysis of consecutive BSG segments to improve retention and monitoring coverage while preserving BP estimation accuracy. Evaluation will combine BP errors, subject-level retention, and effective monitoring duration across heart-rate conditions, supported by additional synchronized recordings during bradycardia.

\section{Conclusions}\label{sec:conclusion}
This paper proposes Phy-BP, a physics-constrained deep learning framework for contactless BP monitoring using triaxial BSG signals. To improve robustness in realistic scenarios, Phy-BP integrates adaptive quality control, triaxial sensing and physics-guided regularization to extract reliable cardiogenic features and align multi-axis representations. Experiments on data from $21$ patients show that Phy-BP outperforms representative PPG/ECG-based and BCG-based comparison models, achieving MAEs of $6.03$, $4.23$ and $5.07$~mmHg for SBP, DBP and MAP, respectively, with a $32.76\%$ reduction in MAP MAE relative to the state-of-the-art model. In the future, we will further investigate whether TPR-related hemodynamic conditions can be inferred by fusing multi-axis BSG representations, enabling calibration-free BP monitoring.

\patchcmd{\IEEEbiographynophoto}{\vskip 4\baselineskip plus 1fil minus 0\baselineskip}{\vskip 0.5\baselineskip}{}{\PackageError{manuscript}{Biography spacing patch failed}{Check the IEEEtran biography definition.}}

\begin{IEEEbiographynophoto}{Yuanyuan Zhang}
is currently a Postdoctoral Associate with the SensorWeb Laboratory at the University of Georgia, Athens, GA, USA. He received the Ph.D. and B.Eng. degrees in Electrical and Electronic Engineering from the University of Liverpool, UK, in 2025 and 2020, respectively, and the M.S. degree in Control Systems from Imperial College London, UK, in 2021. His current research interests include wireless sensing, bioradar systems, bodyseismography, blood pressure monitoring, and physiological digital twins.
\end{IEEEbiographynophoto}

\begin{IEEEbiographynophoto}{Yida Zhang}
is currently pursuing the Ph.D. degree in Electrical and Computer Engineering at the University of Georgia. He received his bachelor's degree in Computer Science and Technology from Hohai University, Nanjing, China, in 2024. His current research interests include mobile computing, human sensing, and physiological signal processing.
\end{IEEEbiographynophoto}

\begin{IEEEbiographynophoto}{Jiahui Li}
is a Ph.D. student in Computer Science at the University of Georgia. His research interests include mobile and wearable sensing, human-centered AI, and healthcare applications.
\end{IEEEbiographynophoto}

\begin{IEEEbiographynophoto}{Yuyan Wu}
is a Ph.D. candidate in Civil and Environmental Engineering at Stanford University. She received her B.S. degree in Applied Physics, with a minor in Computer Science, from the University of Science and Technology of China. Her research focuses on indirect human and animal monitoring through ambient vibration sensing.
\end{IEEEbiographynophoto}

\begin{IEEEbiographynophoto}{Fei Dou}
is an Assistant Professor in the School of Computing at the University of Georgia. She received her Ph.D. in Computer Science and Engineering from the University of Connecticut in 2023. Her research interests include machine learning, ubiquitous computing, intelligent sensing, and AI for health. Fei is the recipient of the Women of Innovation (WOI) Academic Innovation and Leadership Award in Connecticut.
\end{IEEEbiographynophoto}

\begin{IEEEbiographynophoto}{Xiao Yin}
is currently pursuing the Ph.D. degree in Electronic Information at the School of Electronic Science and Engineering, Nanjing University, Nanjing, China. He is also an Attending Physician with the Department of Critical Care Medicine, Yixing People's Hospital, Yixing, China.
\end{IEEEbiographynophoto}

\begin{IEEEbiographynophoto}{Zhenlin An}
received the B.E. degree from the School of Information and Communication Engineering, Dalian University of Technology, China, in 2017, and the Ph.D. degree from the Department of Computing, The Hong Kong Polytechnic University, in 2021. He was a postdoctoral fellow with Princeton University and the University of Pittsburgh. He is currently an Assistant Professor with the School of Computing, University of Georgia. His research interests include wireless communication, indoor localization, and mobile computing.
\end{IEEEbiographynophoto}

\begin{IEEEbiographynophoto}{Hae Young Noh}
is a Professor in Civil and Environmental Engineering (CEE) at Stanford University. She received her Ph.D. and M.S. degrees in CEE and EE from Stanford, and her B.S. degree in Mechanical \& Aerospace Engineering from Cornell. Her research focuses on indirect-sensing and physics-guided data analytics using ``structures as sensors'' to enable scalable, non-intrusive monitoring of cyber-physical-human systems.
\end{IEEEbiographynophoto}

\begin{IEEEbiographynophoto}{WenZhan Song}
(Senior Member, IEEE) received the B.S. and M.S. degrees in computer science from Nanjing University of Science and Technology, Nanjing, China, in 1997 and 1999, respectively, and the Ph.D. degree in computer science from Illinois Institute of Technology, Chicago, IL, USA, in 2005. He is currently the Chair Professor of electrical and computer engineering at the University of Georgia, Athens, GA, USA. He is also the founder of Intelligent Dots LLC, Peachtree Corners, GA, USA. His laboratory invented a series of noninvasive sensing technologies for the health and security monitoring of humans, animals, machines, and infrastructure, driven by the convergence of AI, data science, and advanced sensing and networking technologies. His research focuses on sensor data analytics and security, sensor networks, and data infrastructures.
\end{IEEEbiographynophoto}


\begin{thebibliography}{100}

\bibitem{huai2024bp} K.~Huai, Z.~Zhou, H.~Sun, and X.~Zhang, ``{BP-STFNet}: A hybrid time-frequency domain neural network for blood pressure estimation from multi-channel {BCG} signals,'' in \emph{2024 International Joint Conference on Neural Networks (IJCNN)}.\hskip 1em plus 0.5em minus 0.4em\relax IEEE, Sep. 2024, pp. 1--8.

\bibitem{huang2024ai} Y.~Huang, L.~Chen, C.~Li, J.~Peng, Q.~Hu, Y.~Sun, H.~Ren, W.~Lyu, W.~Jin, J.~Tian \emph{et~al.}, ``{AI}-driven system for non-contact continuous nocturnal blood pressure monitoring using fiber optic ballistocardiography,'' \emph{Communications Engineering}, vol.~3, no.~1, p. 183, Dec. 2024.

\bibitem{litton2012picco} E.~Litton and M.~Morgan, ``The {PiCCO} monitor: A review,'' \emph{Anaesthesia and Intensive Care}, vol.~40, no.~3, pp. 393--408, May 2012.

\bibitem{zhang2025blood} Y.~Zhang, J.~Chen, Y.~Song, Z.~Zeng, X.~Zhang, Q.~Lu, B.~G. Phillips, Z.~Xie, and W.~Song, ``Blood pressure estimation from vibration signals via coarse-to-fine contrastive learning, feature selection and synthesis,'' in \emph{Proceedings of the ACM/IEEE International Conference on Connected Health: Applications, Systems and Engineering Technologies}, Jun. 2025, pp. 134--138.

\bibitem{li2025practicalbp} C.~Li, J.~Peng, Q.~Hu, L.~Chen, Y.~Huang, S.~Wu, W.~Cheng, and Q.~Zhang, ``{PracticalBP}: Continuous cuffless blood pressure monitoring with only one record for calibration,'' \emph{Proceedings of the ACM on Interactive, Mobile, Wearable and Ubiquitous Technologies}, vol.~9, no.~3, pp. 1--31, Sep. 2025.

\bibitem{shahrbabak2025silico} S.~M. Shahrbabak, A.~Mousavi, R.~Mukkamala, and J.-O. Hahn, ``In silico investigation of a mathematical model relating the ballistocardiogram to aortic blood pressure,'' \emph{IEEE Transactions on Biomedical Engineering}, Jul. 2025.

\bibitem{chen2026intrinsic} S.~Chen, H.~Luo, Z.~Yao, Z.~Jiang, X.~Wu, and H.~Liu, ``Intrinsic {PPG-ECG} coupling for accurate and low-power blood pressure monitoring,'' \emph{Advanced Science}, p. e20101, Mar. 2026.

\bibitem{wang2025pitn} R.~Wang, M.~Qi, Y.~Shao, A.~Zhou, and H.~Ma, ``{PITN}: Physics-informed temporal networks for cuffless blood pressure estimation,'' \emph{IEEE Transactions on Mobile Computing}, Apr. 2025.

\bibitem{wang2024accurate} L.~Wang, X.~Wang, Y.~Zhang, X.~Ma, H.~Dai, Y.~Zhang, Z.~Li, and T.~Gu, ``Accurate blood pressure measurement using smartphone's built-in accelerometer,'' \emph{Proceedings of the ACM on Interactive, Mobile, Wearable and Ubiquitous Technologies}, vol.~8, no.~2, pp. 1--28, May 2024.

\bibitem{chang2019cuff} E.~Chang, C.-K. Cheng, A.~Gupta, P.-H. Hsu, P.-Y. Hsu, H.-L. Liu, A.~Moffitt, A.~Ren, I.~Tsaur, and S.~Wang, ``Cuff-less blood pressure monitoring with a 3-axis accelerometer.'' in \emph{2019 41st Annual International Conference of the IEEE Engineering in Medicine and Biology Society (EMBC)}, Jul. 2019, pp. 6834--6837.

\bibitem{wu2025ubicon} Y.~Wu, S.~Bai, Q.~Hu, B.~Wang, M.~Li, X.~Hu, and Y.~Chen, ``{Ubicon-BP}: Towards ubiquitous, contactless blood pressure detection using smartphone,'' \emph{IEEE Transactions on Mobile Computing}, Aug. 2025.

\bibitem{zhang2023overview} Y.~Zhang, R.~Yang, Y.~Yue, E.~G. Lim, and Z.~Wang, ``An overview of algorithms for contactless cardiac feature extraction from radar signals: Advances and challenges,'' \emph{IEEE Transactions on Instrumentation and Measurement}, Aug. 2023.

\bibitem{wu2022facial} B.-F. Wu, B.-J. Wu, B.-R. Tsai, and C.-P. Hsu, ``A facial-image-based blood pressure measurement system without calibration,'' \emph{IEEE Transactions on Instrumentation and Measurement}, vol.~71, pp. 1--13, Apr. 2022.

\bibitem{zhang2024radarODE} Y.~Zhang, R.~Guan, L.~Li, R.~Yang, Y.~Yue, and E.~G. Lim, ``{radarODE: An ODE-embedded deep learning model for contactless ECG reconstruction from millimeter-wave radar},'' \emph{IEEE Transactions on Mobile Computing}, Apr. 2025.

\bibitem{zhang2025high} Y.~Zhang, H.~Zhao, S.~Xiong, R.~Yang, E.~G. Lim, and Y.~Yue, ``From high-{SNR} radar signal to {ECG}: A transfer learning model with cardio-focusing algorithm for scenarios with limited data,'' \emph{IEEE Transactions on Mobile Computing}, Oct. 2025.

\bibitem{cao2025mmwave} Y.~Cao, S.~Zhang, F.~Li, Z.~Chen, and J.~Luo, ``{mmWave-based} contactless {BP} monitoring with physio-model-guided deep learning,'' \emph{IEEE Transactions on Mobile Computing}, Dec. 2025.

\bibitem{shi2025mmbp} Z.~Shi, T.~Gu, Y.~Zhang, and X.~Zhang, ``mmbp+: Contact-free blood pressure measurement using millimeter-wave radar,'' \emph{IEEE Transactions on Mobile Computing}, Jan. 2025.

\bibitem{starr1955studies} I.~Starr, ``Studies made by simulating systole at necropsy. vi. estimation of cardiac stroke volume from the ballistocardiogram,'' \emph{Journal of Applied Physiology}, vol.~8, no.~3, pp. 315--329, May 1955.

\bibitem{starr1959studies} I.~Starr, ``Studies made by simulating systole at necropsy: {XII. Estimation} of the initial cardiac forces from the ballistocardiogram,'' \emph{Circulation}, vol.~20, no.~1, pp. 74--87, Jul. 1959.

\bibitem{kim2018ballistocardiogram} C.-S. Kim, A.~M. Carek, O.~T. Inan, R.~Mukkamala, and J.-O. Hahn, ``Ballistocardiogram-based approach to cuffless blood pressure monitoring: Proof of concept and potential challenges,'' \emph{IEEE Transactions on Biomedical Engineering}, vol.~65, no.~11, pp. 2384--2391, Nov. 2018.

\bibitem{chen2025selfdenoiser} J.~Chen, Y.~Song, Y.~Zhang, Z.~Zeng, X.~Zhang, Z.~Pitafi, Z.~Xie, D.~K. Das, N.~Dong, J.~Lu \emph{et~al.}, ``{SelfDenoiser}: Self-supervised seismic signal denoiser for continuous and contactless cardiac monitoring,'' \emph{Proceedings of the ACM on Interactive, Mobile, Wearable and Ubiquitous Technologies}, vol.~9, no.~4, pp. 1--31, Dec. 2025.

\bibitem{zhang2026bcg} Y.~Zhang, R.~Jiang, Y.~Pan, C.~Wang, G.~Yang, and X.~Zhang, ``{BCG-BP-FSNet}: A few-shot learning framework for personalized, non-invasive blood pressure prediction based on ballistocardiogram,'' \emph{Biomedical Signal Processing and Control}, vol. 120, p. 109891, Jul. 2026.

\bibitem{li2020smart} F.~Li, M.~Valero, J.~Clemente, Z.~Tse, and W.~Song, ``Smart sleep monitoring system via passively sensing human vibration signals,'' \emph{IEEE sensors journal}, vol.~21, no.~13, pp. 14\,466--14\,473, Jul. 2020.

\bibitem{lyu2025contactless} W.~Lyu, T.~Chen, S.~Chen, W.~Yuan, Y.~Li, Q.~Wang, and C.~Yu, ``Contactless arrhythmia detection and short-term {HRV} analysis based on a fiber optic sensor,'' \emph{IEEE Transactions on Instrumentation and Measurement}, Aug. 2025.

\bibitem{xiong2025enhancing} S.~Xiong, C.~Tang, Y.~Zhang, H.~Xiong, Y.~Xu, and A.~Shimada, ``Enhancing nonlinear dependencies of {Mamba} via negative feedback for time series forecasting,'' \emph{Applied Soft Computing}, p. 113758, Dec. 2025.

\bibitem{chu2024vessel} Z.~Chu, R.~Yan, and S.~Wang, ``Vessel turnaround time prediction: A machine learning approach,'' \emph{Ocean \& Coastal Management}, vol. 249, p. 107021, Mar. 2024.

\bibitem{liu2024arm} H.~Liu, D.~Zhao, A.~Sabit, C.~H. Pathiravasan, J.~Ishigami, J.~Charleston, E.~R. Miller~III, K.~Matsushita, L.~J. Appel, and T.~M. Brady, ``Arm position and blood pressure readings: the arms crossover randomized clinical trial,'' \emph{JAMA internal medicine}, vol. 184, no.~12, pp. 1436--1442, Oct. 2024.

\bibitem{kallioinen2017sources} N.~Kallioinen, A.~Hill, M.~S. Horswill, H.~E. Ward, and M.~O. Watson, ``Sources of inaccuracy in the measurement of adult patients’ resting blood pressure in clinical settings: a systematic review,'' \emph{Journal of hypertension}, vol.~35, no.~3, pp. 421--441, Mar. 2017.

\bibitem{wu2026noncontact} Y.~Wu and H.~Y. Noh, ``Non-contact mass estimation of static objects on {Kirchhoff--Love} plates via active vibration sensing,'' Available at SSRN, Jan. 2026, SSRN Scholarly Paper No. 6602436.

\bibitem{song2025multi} Y.~Song, H.~Xiang, Z.~Zeng, J.~Chen, Y.~Zhang, Z.~F. Pitafi, H.~Yang, Q.~Lu, X.~Zhang, B.~G. Phillips \emph{et~al.}, ``Multi-granularity supervised contrastive learning with online adaptation for contactless in-bed posture classification,'' \emph{Proceedings of the ACM on Interactive, Mobile, Wearable and Ubiquitous Technologies}, vol.~9, no.~2, pp. 1--32, Jun. 2025.

\bibitem{bose2024invariance} R.~Bose and A.~M. Roy, ``Invariance embedded physics-infused deep neural network-based sub-grid scale models for turbulent flows,'' \emph{Engineering Applications of Artificial Intelligence}, vol.~128, p.~107483, Feb. 2024.

\bibitem{roy2023elasticity} A.~M. Roy, R.~Bose, V.~Sundararaghavan, and R.~Arr\'{o}yave, ``Deep learning-accelerated computational framework based on Physics Informed Neural Network for the solution of linear elasticity,'' \emph{Neural Networks}, vol.~162, pp.~472--489, May 2023.

\bibitem{starr1950standardization} I.~Starr, O.~Horwitz, R.~Mayock, and E.~Krumbhaar, ``Standardization of the ballistocardiogram by simulation of the heart's function at necropsy; with a clinical method for the estimation of cardiac strength and normal standards for it,'' \emph{Circulation}, vol.~1, no.~5, pp. 1073--1096, May 1950.

\bibitem{Carcione2022} J.~M. Carcione, ``Chapter 2: Viscoelasticity and wave propagation,'' in \emph{Wave Fields in Real Media}, 4th~ed., J.~M. Carcione, Ed.\hskip 1em plus 0.5em minus 0.4em\relax Elsevier, 2022, pp. 63--133.

\bibitem{luporini2020architecture} F.~Luporini, M.~Louboutin, M.~Lange, N.~Kukreja, P.~Witte, J.~H{\"u}ckelheim, C.~Yount, P.~H. Kelly, F.~J. Herrmann, and G.~J. Gorman, ``Architecture and performance of {Devito}, a system for automated stencil computation,'' \emph{ACM Transactions on Mathematical Software (TOMS)}, vol.~46, no.~1, pp. 1--28, Apr. 2020.

\bibitem{cheng2026information} X.~Cheng, W.~Yuan, Y.~Yang, Y.~Zhang, S.~Cheng, Y.~He, and Z.~Sun, ``Information Shapes Koopman Representation,'' in \emph{International Conference on Learning Representations}, 2026.

\bibitem{xiong2026kuma} S.~Xiong, C.~Tang, and A.~Shimada, ``KUMA: A Novel Framework with Koopman Separation and Efficient Multilevel Extraction in Time Series Forecasting,'' in \emph{International Conference on Machine Learning}, 2026.

\bibitem{he2016identity} K.~He, X.~Zhang, S.~Ren, and J.~Sun, ``Identity mappings in deep residual networks,'' in \emph{European conference on computer vision}.\hskip 1em plus 0.5em minus 0.4em\relax Springer, Sep. 2016, pp. 630--645.

\bibitem{chen2026cows} X.~Chen, R.~Yang, Y.~Xue, and Z.~Wang, ``{CoWS}: Self-supervised representation pre-training for cross-machine fault diagnosis,'' \emph{IEEE Transactions on Emerging Topics in Computational Intelligence}, vol.~10, no.~2, pp.~1311--1323, Apr. 2026.

\bibitem{boubekki2026explaining} A.~Boubekki, K.~Patlatzoglou, J.~Barker, H.~Aggour, F.~S. Ng, and A.~H. Ribeiro, ``Explaining deep learning for ECG using time-localized clusters,'' \emph{IEEE Transactions on Biomedical Engineering}, early access, pp.~1--19, Jun. 2026.

\bibitem{rgi45hz} {RacoTech}, ``{RGI-4.5Hz} low frequency geophone element,'' 2018, \url{http://www.racotech.biz/product.php?id=84}, Accessed: Mar. 8, 2025.

\bibitem{song2026sleepapnea} Y.~Song, J.~Chen, Z.~Zeng, Y.~Zhang, Z.~Pitafi, D.~K. Das, B.~G. Phillips, Y.~Kwon, W.~J. Healy, X.~Zhang \emph{et~al.}, ``Contactless Sleep Apnea Detection with Bodyseismography,'' \emph{Proceedings of the ACM on Interactive, Mobile, Wearable and Ubiquitous Technologies}, vol.~10, no.~2, pp.~1--31, Jun. 2026.

\bibitem{huang2026hyde} Y.~Huang, C.~Li, C.~Li, L.~Chen, J.~Peng, W.~Cheng, Q.~Zhang, and K.~Wu, ``{Hyde}: Inferring cardiovascular hemodynamics using {BCG} for cross-user hypertension detection,'' \emph{IEEE Transactions on Mobile Computing}, vol.~25, no.~10, pp.~17\,134--17\,149, Oct. 2026.

\bibitem{fan2023fewshotbp} F.~Fan, Y.~Gu, J.~Shen, F.~Dong, and Y.~Chen, ``{FewshotBP}: Towards personalized ubiquitous continuous blood pressure measurement,'' \emph{Proceedings of the ACM on Interactive, Mobile, Wearable and Ubiquitous Technologies}, vol.~7, no.~3, pp. 1--39, Sep. 2023.

\bibitem{kim2016mechanism} C.-S.~Kim, S.~L. Ober, M.~S. McMurtry, B.~A. Finegan, O.~T. Inan, R.~Mukkamala, and J.-O.~Hahn, ``Ballistocardiogram: Mechanism and potential for unobtrusive cardiovascular health monitoring,'' \emph{Scientific Reports}, vol.~6, no.~1, p.~31297, Aug. 2016.

\bibitem{zhan2024cardiac} J.~Zhan, X.~Wu, X.~Fu, C.~Li, K.-Q.~Deng, Q.~Wei, C.~Zhang, T.~Zhao, C.~Li, L.~Huang \emph{et~al.}, ``Non-contact assessment of cardiac physiology using FO-MVSS-based ballistocardiography: A promising approach for heart failure evaluation,'' \emph{Scientific Reports}, vol.~14, no.~1, p.~3269, Feb. 2024.

\bibitem{agam2022diastolic} A.~Agam, P.~S\o gaard, K.~Kragholm, A.~S. Jensen, K.~S\o rensen, J.~Hansen, and S.~Schmidt, ``Correlation between diastolic seismocardiography variables and echocardiography variables,'' \emph{European Heart Journal -- Digital Health}, vol.~3, no.~3, pp.~465--472, Sep. 2022.

\end{thebibliography}
\end{document}